\documentclass[a4paper,11pt]{article}
\pdfoutput=1 % if your are submitting a pdflatex (i.e. if you have
\usepackage{jcappub} % for details on the use of the package, please
\usepackage[T1]{fontenc} % if needed

\usepackage{bm}
\usepackage{latexsym}
\usepackage{amsmath,amsfonts,amssymb}
\usepackage{graphicx,epsfig}
\usepackage{psfrag}
\usepackage{amsthm}
\usepackage{color}
\usepackage{braket}
\usepackage{indent first}
\usepackage[normalem]{ulem}
\usepackage{fancyhdr}
\usepackage{subcaption}
\usepackage{hyperref}
\usepackage[titletoc,title]{appendix}
 \usepackage{environ}
\usepackage{mathtools}
\usepackage{float}
\usepackage[makeroom]{cancel}
\def\bea{\begin{equation}}
\def\eea{\end{equation}}
\def\beq{\begin{eqnarray}}
\def\eeq{\end{eqnarray}}
\title{Matter-antimatter asymmetry in generalized coupling theories}
\author[a,c,1]{A. Troisi,\note{Corresponding author.}}
\author[b,c]{G. Lambiase,}
\author[d,e,f]{S. Carloni,}

\affiliation[a]{Dipartimento di Scienze e Tecnologie, Universit\`{a} del Sannio, Via De Sanctis, Benevento I-82100, Italy}
\affiliation[b]{Diparatimento di Fisica E.R. Caianiello, Universit\'a di Salerno, Italy}
\affiliation[c]{INFN, Sezione di Napoli Gruppo Collegato di Salerno, Complesso Universitario di Monte S. Angelo, I-80126 Napoli, Italy}
\affiliation[d]{DIME Sez. Metodi e Modelli Matematici, Universit\`{a} di Genova, Via  All’Opera Pia 15, 16145 - Genoa, (Italy).}
\affiliation[e]{INFN Sezione di Genova, Via Dodecaneso 33, 16146 Genova, Italy}
\affiliation[f]{Institute of Theoretical Physics, Faculty of Mathematics and Physics,
Charles University, Prague, V Hole{\v s}ovi{\v c}k{\' a}ch 2, 180 00 Prague 8, Czech Republic}

\emailAdd{antro@unisannio.it}
\emailAdd{lambiase@sa.infn.it}
\emailAdd{sante.carloni@unige.it}
\abstract{
We explore the gravitational baryogenesis paradigm in the homogeneous and isotropic cosmology of generalized coupling gravity and, in particular, of the so-called Minimal Exponential Measure Model (MEMe). We show that, also in this theory, the time derivative of the Ricci scalar couples with matter currents and can preserve an unbalance in the baryon-antibaryon number
beyond thermal equilibrium. Using the current bounds on the ratio of baryon number to entropy density, we can considerably improve the known constraints on the parameter $q$ that characterizes the MEMe model. This estimate also allows us to draw stringent constraints on the spatial curvature of the cosmological model.}

\begin{document}
\maketitle
%
%%%%%%%%%%%%%%%%%%%%%%%%%%%%%%%%%%%%%%%%%%%%%%%%%%%%%%
\section{Introduction}
%%%%%%%%%%%%%%%%%%%%%%%%%%%%%%%%%%%%%%%%%%%%%%%%%%%%%%

In a series of recent papers \cite{Carloni:2016glo,FengCarloni2019,FengCarloni2021,Feng:2022rga}, a new class of modifications of General Relativity (GR), called ``generalized coupling theories'', was proposed and analyzed. These new theories are constructed not by assuming that the gravitational interaction departs from General Relativity at a specific scale but by postulating that the bending of spacetime is described by a non-linear relation between the Einstein tensor and the stress-energy tensor \cite{Carloni:2016glo}. In this way, the successful results of GR, which are related to vacuum solutions (e.g., gravitational waves), can be preserved, but there is still space for new physics, which could provide a framework for unresolved issues in relativistic gravitation.  

One of the simplest possible realizations of a generalized coupling theory, the so-called Minimal Exponential Measure Model (MEMe), has the advantage of depending on just one additional parameter,  $q$, with respect to GR. Yet, it is able to explain the onset of a dark energy era. The MEMe model has been analyzed in different contexts, and, using data coming from gravitational waves \cite{Carloni:2016glo} and predictions from the Post-Newtonian analysis \cite{FengCarloni2021}, it has been possible to derive an upper bound for $q$. Indeed, since the value of $q$ is crucial to determining the characteristic scales at which departures from GR arise, it becomes necessary to estimate as accurately as possible the value of such a parameter. Some of the tightest known constraints on gravitational theories are known to come from the analysis of nuclear and subnuclear processes in the early universe. Hence, it is natural to turn to this sector of cosmology to refine our estimation of $q$.

In this paper, we focus on bounds that stem from observational data related to the origin of baryon asymmetry in the Universe \cite{kolb}. As well known, the predictions of the Big Bang Nucleosynthesis \cite{Copi,burles} and the observations of Cosmic Microwave Background anisotropies combined with the large structure of the Universe \cite{wmap,bennet} provide an estimate of the parameter  characterizing such an asymmetry given by  \cite{AharonyShapira:2021ize,Ade}
 \begin{equation*}\label{etaexp}
\eta \equiv \frac{n_B-n_{\bar B}}{s}\lesssim (
9.2 \pm 0.5)\,\, 10^{-11}\,,
 \end{equation*}
where $n_{B}/n_{\bar B}$ denote the number densities of baryon/antibaryon in the Universe, 
$$s=\frac{2\pi^2}{45}g_{*s}{\mathcal T}^3,$$ 
the entropy density of the Universe $g_{*s}\simeq 106$ counts the total degrees of freedom for particles that contribute to the entropy of the Universe and ${\mathcal T}$ is its temperature.

Baryon asymmetry represents a puzzling conundrum that, today, has been modeled following several approaches \cite{Kolb:1983ni,Shaposhnikov:1987tw,Sakharov:1988vdp}. A suitable interpretative scheme has been provided by Sakharov, who described three basic conditions for a CPT invariant theory to explain the baryon asymmetry \cite{sakharov}:
1)  processes exist that violate the baryon number; 2) the discrete C and CP symmetries are violated (C is the charge conjugation, CP the combination of the charge conjugation and the parity; 3) thermal equilibrium is broken.
Sakharov's conditions may be relaxed in some cases \cite{dolgov}. In particular, if the CPT symmetry is violated dynamically \cite{cohen}, the generation of a net baryon number asymmetry can also be allowed, preserving thermal equilibrium. This mechanism is the essential ingredient of the gravitational baryogenesis model \cite{gravbar}, where the baryon/lepton current is coupled to the derivative of the scalar curvature of spacetime, providing a straightforward asymmetry mechanism in the matter/antimatter sector. As we will discuss in the following, such a coupling is naturally nonvanishing in the framework of the MEMe model. As a matter of fact, we can investigate the necessary conditions for generating a net baryon asymmetry in the universe. Exploiting baryon asymmetry observational results, we get an upper bound on the parameter $q$ that characterizes the MEMe cosmological model.

The paper is organized as follows. In Section II, we
briefly resume the conceptual bases of the MEMe model. 
In Section III, we discuss the origin of matter-antimatter asymmetry in the framework of the gravitational baryogenesis model. In Sec. IV, we implement the gravitational baryogenesis formalism in the framework of Minimal Exponential Measure gravity, and we infer the upper bound on the $q$ parameter.  In Sect. V is devoted to a general analysis of the results and to drawing our conclusions. 

%%%%%%%%%%%%%%%%%%%%%%%%%%%%%%%%%%%%%%%%%%%%%%%%%%%%

\section{The Minimal Exponential Measure Model}

%%%%%%%%%%%%%%%%%%%%%%%%%%%%%%%%%%%%%%%%%%%%%%%%%%%%

Generalized coupling theories are characterized by the field equations \cite{Carloni:2016glo}
\begin{equation}\label{GCA-EFEgen}
G_{\mu \nu} = \chi_{\mu \nu}{^{\alpha \beta}} \, T_{\alpha \beta},
\end{equation}
derived by the action principle 
\begin{equation}\label{GCA-Action}
S= \int d^4x \biggl\{ \frac{1}{2 \kappa} R\sqrt{-{g}} 
+L[\phi,\mathfrak{g}]\sqrt{-\mathfrak{g}} \biggr\}.
\end{equation}
where $G_{\mu \nu}$ is the Einstein tensor, $\kappa = 8 \pi G$, $L[\phi,\mathfrak{g}]$ is the Lagrangian density associated with other fields of the theory, including standard matter. The metrics ${g}_{\mu \nu}$  and $\mathfrak{g}_{\mu \nu}$ are related by means of an auxiliary field $\chi_{\mu \nu}{^{\alpha \beta}}$ in the following way
\begin{equation}\label{Rel2g}
\mathfrak{g}_{\mu \nu} = \chi_{\mu \nu}{^{\alpha \beta}} \, g_{\alpha \beta}.
\end{equation}
Assuming that in the absence of matter the coupling tensor reduces to 
\begin{equation}
   \left. \chi_{\mu \nu}{^{\alpha \beta}}\right|_{vacuum}=\delta_{\mu}^{\alpha}\delta_{\nu}^{\beta}\,,
\end{equation} 
the vacuum solutions of the new theory exactly match ones of General Relativity. As a consequence, this class of theories provides the same phenomenology of GR in vacuum and passes all the astrophysical and Solar System tests. However, when matter cannot be neglected, i.e., in cosmology or the interior of relativistic stars, a significant difference with respect to Einstein's theory can occur. 

In general, one can write the field equations in terms of the metric $g$ or $\mathfrak{g}$, and so the choice of description is akin to the choice of field definition in a conformal transformation. As such, we name, abusing the terminology in favor of a simpler reference, $g$ as the {\it Einstein metric} as the action written in terms of this metric presents nonminimal coupling terms, and $\mathfrak{g}$ the {\it Jordan metric}. Consequently, the field equations obtained by \eqref{GCA-Action} written in terms of $g$ will be indicated by the term {\it Einstein frame}, whereas the ones written in terms of $\mathfrak{g}$ will represent the {\it Jordan frame}.

The \textit{Minimal Exponential Measure} (MEMe) model is a particular type of generalized coupling model in which the coupling tensor is written in terms of a non-kinetic rank two tensor  $A{_\mu}{^\alpha}$  as (see \cite{FengCarloni2019} for more details)
\begin{equation}\label{GCA-JordanMetric}
\chi_{\mu \nu}{^{\alpha \beta}}  = \,  A{_\mu}{^\alpha} \, A{_\nu}{^\beta} .
\end{equation}
The action for  $A{_\mu}{^\alpha}$  is chosen to be 
\begin{equation}
S_A=- \int d^4x \frac{\lambda}{\kappa} \sqrt{-\mathfrak{g}}
\end{equation}
where $\lambda$ is a coupling constant. Hence, within this scheme, the total general action can be written as \cite{FengCarloni2021}
\begin{equation}\label{GCA-MEMeAction}
S[\phi,g,A{_{\cdot}}{^{\cdot}}]= \int d^4x \biggl\{ \frac{1}{2 \kappa}\left[R - 2(\Lambda -\lambda)\right]\sqrt{-{g}} 
+ \left(L_{m}[\phi,\mathfrak{g}] - \frac{\lambda}{\kappa} \right) \sqrt{-\mathfrak{g}} \biggr\} ,
\end{equation}where $\phi$ represents a generic matter field.  In this model, an effective cosmological term $\Lambda_{eff}$ is present: 
\begin{equation}\label{LambdaeffG}
    \Lambda_{eff}=\Lambda-\lambda\left(1- \frac{\sqrt{\mathfrak{g}}}{\sqrt{g}}\right)
\end{equation}
which represents a cosmological constant term $\Lambda$ when $\mathfrak{g}\approx {g}$, but it is in general non-constant due to the last term in (\ref{LambdaeffG}). 

As no kinetic term for $A{_\mu}{^\alpha}$ appears in the action, this coupling field is non-dynamical, thus evading the no-go result of \cite{Pani:2013qfa}. With these choices, relation \eqref{Rel2g} can be written as 
\begin{equation}\label{GCA-JordanMetric}
\mathfrak{g}_{\mu \nu} = A{_\mu}{^\alpha} \, A{_\nu}{^\beta} \, g_{\alpha \beta} .
\end{equation}
which also implies
\begin{equation}\label{det_g_goth}
\sqrt{-\mathfrak{g}}=\sqrt{-g}|A|.
\end{equation}
The field equations for the coupling tensor ${A}{_\beta}{^\alpha}$ in the Einstein frame can be put in the form
\begin{equation}\label{GCA-ExpFEs}
\begin{aligned}
{A}{_\beta}{^\alpha} - \delta{_\beta}{^\alpha} = q \left[ \frac{1}{4} \mathfrak{T} \, {A}{_\beta}{^\alpha} - \mathfrak{T}_{\beta \nu} \, \mathfrak{g}^{\alpha \nu} \right].
\end{aligned}
\end{equation}
where $ \mathfrak{T}_{\mu \nu}$ is the  Jordan frame stress-energy tensor obtained by the variation of $L_{m}$ with respect to $\mathfrak{g}_{\mu \nu}$ and we have set $q=\kappa\lambda^{-1}$ for later convenience. 

The possibility of considering two different metrics automatically selects two natural frames to describe gravitational phenomenology: the free-falling frame for the metric $g$, described by the timelike vector $U^a$, and the one for $\mathfrak{g}$, described by the timelike vector $u^a$. As $\mathfrak{g}$ is the metric effectively perceived by the matter fields, we can conclude that $u^a$ can always be chosen to represent a frame comoving with matter.  In this case, if we also assume matter to be a perfect fluid in its rest frame, which is free-falling, we can set 
\begin{equation} \label{GCA-EnergyMomentumPerfectFluid}
\mathfrak{T}_{\mu \nu} = \left(\rho + p\right)u_\mu u_\nu + p \> \mathfrak{g}_{\mu \nu},
\end{equation}
and we can express  $A{_\mu}{^\alpha}$ in terms of the matter variables. In particular, we have 
\begin{equation}\label{GCA-Ansatz}
A{_\beta}{^\alpha} = {Y} \, \delta{_\beta}{^\alpha}  + {Z} \, u{_\beta} \, u{^\alpha} ,
\end{equation}
where
\begin{equation}\label{GCA-ExpYsoln}
Y = \frac{4 (1 - p \, q)}{4 - q \,  (3 \, p - \rho)},
\end{equation}
\begin{equation}\label{GCA-ExpZsoln}
Z = - \frac{q \, (p + \rho) [4 - q \, (3 \, p - \rho)]}{4 \, (q \, \rho + 1)^2}.
\end{equation}
Notice that, effectively, the form that we have found for the coupling tensor corresponds to a disformal transformation between $\mathfrak{g}_{\mu \nu}$ and ${g}_{\mu \nu}$. In addition, since, by definition, we have
\begin{equation} \label{GCA-UnitNorm1}
U^\mu \, U^\nu \,{g}_{\mu \nu} = -1, 
\end{equation}
and 
\begin{equation} \label{GCA-UnitNorm2}
u^\mu \, u^\nu \, \mathfrak{g}_{\mu \nu} = -1,
\end{equation}
we can write 
\begin{equation}\label{GCA-UnitFlowField}
U^\mu = {u^\mu}/{\sqrt{-\varepsilon}} ,
\end{equation}
with 
\begin{equation}\label{GCA-Expepssoln}
\varepsilon = -\frac{16 \, (q \, \rho + 1)^2}{[4 - q \, (3 \, p - \rho)]^2}.
\end{equation}
Thus, the free-falling observers in the Einstein frame have velocities parallel with respect to the ones of the Jordan frame.   Taking into account the above relations, we can now write the MEMe field equations in an Einstein-like form 
\begin{equation}\label{EFQMEMe}
G_{\mu \nu}=\kappa \, T_{\mu \nu},
\end{equation}
where $T_{\mu \nu}$ is the effective energy-momentum tensor defined by
\begin{equation}\label{GCA-ExpTmnEffDecomp}
T_{\mu \nu} = T_1 \, U_\mu \, U_\nu + T_2 \, g_{\mu \nu} ,
\end{equation}
with
\begin{equation}\label{GCA-ExpTmnEffDecompTs}
T_1  = |A| \, (p + \rho), \quad
T_2  = \frac{|A| \, (p \, q - 1) + 1}{q}-\frac{\Lambda}{\kappa},
\end{equation}
and $|A|$ is the determinant given by
\begin{equation}\label{GCA-ExpAdet}
|A|=\frac{256 \, (1 - p \, q)^3 (q \, \rho + 1)}{[4 - q \, (3 p - \rho) ]^4}.
\end{equation}
Some remarks are now in order. In this framework, the field theory that represents matter is constructed with $\mathfrak{g}_{\alpha\beta}$ and, considering  the above results, Eq.~\eqref{GCA-JordanMetric}  can be written as
\begin{equation}\label{GCA-TransformedJordanMetric}
\mathfrak{g}_{\mu \nu} =  Y^2 \, g_{\mu \nu} -\varepsilon Z \, (2 \, Y + \varepsilon \, Z) \, U_\mu \, U_\nu.
\end{equation} 
Because of the form of $Y$,$Z$ and $\epsilon$, one can notice that Eq.\eqref{GCA-TransformedJordanMetric} leads to
$\mathfrak{g}_{\mu \nu} =0$ if  $q p=1$ or $q\rho=-1$. Since $\mathfrak{g}_{\mu \nu}$ is effectively singular for these energy density and pressure values, a relativistic quantum field theory of matter based on this metric will also fail. Thus, $q=\kappa/\lambda$ can be considered as a critical scale at which a field description of matter based on $\mathfrak{g}$ breaks down, and this means that the MEMe model determines a natural regularization scale for quantum fields, which differs from the Planck scale. These features suggest that one can imagine the MEMe model as a (toy) coarse-grained representation of quantum field theory in curved spacetime.

Eqs. \eqref{EFQMEMe} can be used to construct and analyze cosmologies based on MEMe models. In exploring such models, one has to pay particular attention to the choice of the {\it fundamental observers}. A choice that is in line with the GR approach would be to select observers comoving with matter (i.e., in the MEMe model, would coincide with $u^\mu$). However, such a choice has the drawback of being ill-defined when the Jordan frame breaks down. To avoid such a restriction, we can instead use $U^\mu$. As we have seen, these two four vectors are parallel to each other so that they are both still orthogonal to the three-surfaces of homogeneity described by $\mathfrak{h}_{\mu \nu}=\mathfrak{g}_{\mu \nu}+u_\mu \, u_\nu$ and therefore no ``tilting'' effect \cite{King:1972td}
will be present. 

In this picture, one can assume that for a homogenous and isotropic fluid source, the metric of the spacetime in the frame specified by $U^\mu$ has the Friedmann-Lema\^{\i}tre-Robertson-Walker (FLRW) form
\begin{equation}\label{FLRW}
ds^2=-dt^2+S^2\left(t\right)\left[\frac{dr^2}{1-kr^2}+r^2d\Omega^2\right],
\end{equation}
where $k=-1,0,1$ is the spatial curvature, $d\Omega^2$ the infinitesimal solid angle and $S$ is the scale factor. In the (Einstein) $U^\mu$ frame, the cosmological equations can be written as
\begin{align}
 3q\left(H^2 +\frac{k}{S^2}\right)=& \frac{256 \kappa (1-pq)^3 (q \rho +1)^2}{[4+q (\rho-3p)]^4}+q\Lambda -\kappa,\label{FriedEq}\\
6q\left(\dot{H}+H^2\right)=& -\frac{256\kappa (p q-1)^3 (q \rho +1) [2-q(\rho+3p)]}{ [4+q (\rho-3p)]^4} +2 (q \Lambda-\kappa),\label{RayEq}
\end{align}
where the expansion rate of the Universe $H$ reads as usual $H=\dot{S}/S$, and we have used a dot to indicate the derivative projected along $U^\mu$\footnote{Notice that in the case $w=1/3$, these equations appear to be inconsistent with General Relativity in the limit $q\rightarrow\infty$. Indeed, for this value of the barotropic index, it seems that the first term on the R.H.S of \eqref{FriedEq} diverges. However, such divergence is compensated by a similar behavior of the time derivative $ U^\mu(q)\partial_\mu$ (see \eqref{GCA-UnitFlowField}  and \eqref{GCA-Expepssoln}). When this fact is recognized, the equations can be realized as consistent. This aspect is important for the following discussion as the expression of the baryon asymmetry parameter, which we calculate in $w=1/3$, will appear to have the same pathology.}. If we consider a barotropic equation of state for the fluid, $p=w\rho$, where $w$ is the adiabatic index, from the conservation of the energy-momentum tensor $\nabla^\alpha T_{\alpha \beta}=0$, we can easily obtain the relation
\begin{equation}\label{ConsLaw}
\dot{\rho}=-\frac{3 H \rho  (w+1) \left[q^2 \rho ^2 w (3 w-1)+\rho  (q-7 q w)+4\right]}{q^2
   \rho ^2 w (3 w-1)-q \rho  \left(3 w^2+13 w+2\right)+4}.
\end{equation}
As in GR, the three equations (\ref{FriedEq}-\ref{ConsLaw}) are redundant. For the following discussion, it will also be useful to derive the conservation law in the Jordan frame. Naming $\tilde{\nabla}$ the covariant derivative associated with the Jordan metric, we have \cite{FengCarloni2019}
\begin{equation}\label{ConsJ}
 0=u^{\mu}\mathfrak{g}^{\alpha \nu}\tilde{\nabla}_\alpha\mathfrak{T}_{\mu \nu}= u^{\mu}\tilde{\nabla}_\mu \rho+ \mathcal{H}(\rho+p), \qquad \mathcal{H}=\frac{u^{\mu}\tilde{\nabla}_\mu\mathcal{S}}{\mathcal{S}}
\end{equation}
where
\begin{equation}\label{SEtoSJ}
    \mathcal{S}(t)=Y\,S(t).
\end{equation}

The above considerations make evident the necessity to constrain the value of $q$ using experimental data. Up to now, 
two different phenomena have been considered to achieve this task. The first relies on the natural difference between the electromagnetic and gravitational waves in the MEMe model. In fact, within the MEMe model, electromagnetic waves propagate according to the metric $\mathfrak{g}_{\mu \nu}$, while linearized gravitational waves propagate on a background metric $g_{\mu \nu}$. Considering the constraints on gravitational wave speed coming from the GW170817 kilonova event \cite{LIGOScientific:2017vwq} one obtains  \cite{FengCarloni2019}  $q<0$ and 
\begin{equation}\label{GWqConstraint}
|q| \lesssim  
-2 \times 10^{33}GeV^{-4}.
\end{equation}  
A second study involving the PPN analysis of  circular orbits within spherically symmetric matter distributions allows, to confirm the sign of $q$  and to establish stronger constraints on its modulus \cite{FengCarloni2021}
\begin{equation}\label{GCAqConstraint}
|q| \lesssim 
10^{23} GeV^{-4}.
\end{equation}
Notice that, since the Earth and its telescopes can be considered, in a very good approximation, as comoving with matter, we can assume cosmological observations to be performed in the Jordan frame. Hence, the measurements on the amount of barionic matter in the Universe constitute an estimation of $\rho$, and the constraints on $q$ made above allow us to conclude that at the present time $q\rho\ll 1$. In this regime, the coupling tensor components approach constant values, and the metrics $g$ and $\mathfrak g$ converge, making the Einstein and Jordan frames equivalent. Indeed, we can also infer that the regime $q p,q\rho\sim1$ is only realized in the very early history of the Universe. This result will be particularly relevant for us, as it will allow us to estimate more easily the parameters we need to obtain our results.

In the following sections, we will consider a different approach to constrain $q$. The basic idea is to change the perspective by considering the early universe measurements. To this end, we refer to the matter-antimatter asymmetry estimate after developing a suitable scheme for baryogenesis in the MEMe model. The starting point is the adoption of the gravitational baryogenesis paradigm.

%%%%%%%%%%%%%%%%%%%%%%%%%%%%%%%%%%%%%%%%%%%%%%%%%%%%%%

\section{Gravitational baryogenesis}
\label{Bar}

 %%%%%%%%%%%%%%%%%%%%%%%%%%%%%%%%%%%%%%%%%%%%%%%%%%%%%%
 
Among the several approaches to baryogenesis, supergravity (SUGRA) theories may provide a suitable mechanism for generating a net baryon asymmetry during the first stages of the Universe's evolution \cite{KU1,KU2}. In this scheme, the interaction that induces the (dynamical) CPT violation is given by a coupling between $\partial_\mu R$, that is the derivative of the Ricci scalar curvature $R$, and the baryon/lepton current\footnote{The current ${\cal J}^{\mu}$ may generate a net $B-L$ charge in equilibrium (here $B, L$ are the usual baryon/lepton number) so that the asymmetry is not wiped out by the electroweak anomaly %\cite{krs1985}
.} ${\cal J}^\mu$ \cite{gravbar}
	\begin{eqnarray}\label{intterm}
		\frac{1}{M_*^{2}}\int\mathrm{d}^4x\sqrt{-g}\,{\cal J}^\mu\partial_\mu R\,.
	\end{eqnarray}
In Equation (\ref{intterm}) $M_*$ is the cutoff mass scale that characterizes the effective theory and can be of the order of the Planck mass $M_P\sim 10^{19}$ GeV, or of the order of the upper bound on the tensor mode fluctuation constraints in inflationary scale $M_I\sim 3.3 \times 10^{16}$GeV \cite{gravbar}. As thermal equilibrium is preserved, the third Sakharov condition mentioned in the Introduction can be relaxed, leading to the creation and preservation of the baryon asymmetry.
Some applications of gravitational baryogenesis can be found in Refs.
\cite{Lambiase:2006dq,Lambiase:2006ft,allGB1,allGB2,allGB3,allGB4,allGB5,allGB6,allGB7,allGB8,allGB9,GGLuciano,Das:2021nbq,Lambiase:2013haa}. 
In an expanding universe, the interaction (\ref{intterm}) dynamically breaks $CPT$, generating an asymmetric energy shift between particles and antiparticles. Interactions that violate baryon processes in thermal equilibrium favor the arising of a net baryon asymmetry. Indeed, the latter gets frozen at the decoupling temperature ${\mathcal T}_D$; therefore, the baryon asymmetry remains fixed (this occurs when the expansion rate of the Universe becomes much larger than the interaction rate).

For an expanding universe, whose matter content is described by a perfect fluid, with a four-velocity $v^\mu$, one gets \cite{KU1,KU2}
\begin{equation}\label{Jmu}
 {\cal J}^\mu=(n_B-n_{\Bar{B}})v^\mu\,,
\end{equation}
where $n_B$ and $n_{\Bar{B}}$ represent baryon and anti-baryon number density respectively.
As in standard FLRW spacetimes, one describes gravitational physics from the point of view of a comoving observer, $v^\mu$ can be associated with the four-velocity of such observers.
When the temperature of the Universe drops below the decoupling temperature ${\mathcal T}_D$, the interaction Lagrangian in Eq. (\ref{intterm}) reads \cite{gravbar} 
\begin{eqnarray}\label{integrand}
\frac{1}{M_*^2} {\cal J}^\mu\partial_\mu R=\frac{1}{M_*^2}(n_B-n_{\Bar{B}})\mathring{R}\,,
\end{eqnarray}
here, $\mathring{R}=v^\mu\partial_\mu R$ denotes the time derivative of the Ricci scalar measured by the comoving observers. The above relation allows us to define the effective chemical potential for baryons $\mu_B$ and for antibaryons $\mu_{\Bar{B}}$ as
 $\mu_B=-\mu_{\Bar{B}}=-\displaystyle{\frac{\mathring{R}}{M_*^2}}$ \cite{gravbar}.
Now, for relativistic particles, corresponding to the radiation-dominated era of the Universe, the baryon number density can be calculated by means of the relation \cite{kolb}
\begin{eqnarray}
n_B-n_{\Bar{B}}=\frac{g_b}{6}\mu_B{\mathcal T}^2~,
\end{eqnarray}
where $g_b\sim\mathcal{O}(1)$ is the number of intrinsic degrees of freedom of baryons. Finally, using all these relations, we can write the baryon asymmetry parameter $\eta$  as \cite{kolb}
\begin{eqnarray}\label{base}
\label{asym1}
\eta\equiv\frac{n_B-n_{\Bar{B}}}{s}\simeq-\frac{15\,g_b}{4\pi^2g_*}\frac{\mathring{R}}{M_*^2{\mathcal T}}\bigg|_{{\mathcal T}_D}\,.	
\end{eqnarray}
 Eq. (\ref{base}) indicates that the baryon asymmetry parameter is different from zero provided that $\mathring{R}\neq 0$.

In GR, the time derivative of $R$ can be computed from the trace of the Einstein field equations: $R=-\kappa\, (\rho-3p)$ for a perfect fluid. During the radiation-dominated era, in which we are interested, the adiabatic index of the cosmological fluid is given by $w=1/3$ ($p=\rho/3$), which implies $R=0$. In such a case, $\mathring{R}=0$, and no baryon asymmetry can be generated in the framework of gravitational baryogenesis ($\eta \sim \mathring{R}=0$). The whole perspective changes for cosmological models that allow a non-zero time derivative of the Ricci scalar. As we shall see in the following, this is the case for the MEMe model.

\section{Baryogenesys in the Minimal Exponential Measure model}

Let us now consider the gravitational baryogenesis machinery in the case of the MEMe model. Before starting, there are some important remarks to make. 

First,  some considerations on the origin of the term \eqref{intterm} within our gravitational theory are in order. The MEMe model is fundamentally different from other theories treated in \cite{Lambiase:2006dq,Lambiase:2006ft,allGB1,allGB2,allGB3,allGB4,allGB5,allGB6,allGB7,allGB8,allGB9,GGLuciano,Das:2021nbq,Lambiase:2013haa,Lambiase:2024yfu} since an additional metric $\mathfrak{g}$ is present. Despite the possible similarity between the Einstein and Jordan frames we have mentioned in Section 2, when dealing with baryogenesis, it is unclear which of the two metrics presented in the theory should be considered to calculate the Ricci scalar in \eqref{intterm}. As mentioned above, the arguments traditionally used to motivate terms like \eqref{intterm} arise as part of the effective action approach to quantum gravitational corrections, or also in the context of supergravity. Leaving aside the interpretation via SUGRA theories, which would be too complex and delicate to be presented here, we reason in terms of effective action\footnote{We do not provide here a complete analysis of the structure of the MEMe model as an effective field theory. Although such an analysis would be in principle required to support our arguments and the results we obtain, it is well beyond the scope of this paper and, as we will see, not strictly necessary for our purposes.}. As explained in \cite{Carloni:2016glo}, one might interpret the metric $\mathfrak{g}$ as a toy representation of properties of the matter source that does not fit into the classical perfect fluid description (including, e.g., quantum effects) but that, at the same time, can influence the way in which a fluid gravitates. In this picture, nothing has been said about the quantum nature of spacetime, which is fundamentally represented by $g$. Therefore, one might apply to this metric the classical arguments associated with constructing the effective action in General Relativity. In this way, it is natural to expect a term like \eqref{intterm} to appear when quantum gravitational corrections are considered. These terms will involve the Ricci scalar constructed with ${g}$ only. Based on these considerations, we will assume that the term \eqref{intterm} will contain the Ricci scalar $R(g)$ associated with the metric $g$.

Second, in the definition \eqref{Jmu} of the current ${\cal J}^\mu$, the vector  $v^\mu$ must be identified by definition with  $u^\mu$, as, in our hypothesis, this vector describes the actual matter flow. However, to exploit the cosmological equations (\ref{FriedEq})-(\ref{ConsLaw}), we need to describe the process from the point of view of an observer described by $U^\mu$ (Einstein frame). This fact implies that the time derivative of the Ricci scalar has to be expressed in terms of this vector
\begin{equation}\label{Rdotframe}
\mathring{R}=u^\mu\partial_\mu R= {\sqrt{-\varepsilon}} U^\mu\partial_\mu R={\sqrt{-\varepsilon}}\dot{R}\,.
\end{equation}
Third, unlike the case of GR, it will not be generally possible for us to neglect the contribution of the cosmological constant and the spatial curvature. This is motivated by the fact that the nonlinear source term in Eqs.~\eqref{FriedEq} does not allow, in the Einstein frame, the definition of separated eras of domination of the classical different source term. As the inclusion of spatial curvature complicates the calculation substantially, in the following, we will first assume $k=0$ and then consider the more general case.

\subsection{The spatially flat case}
	
As a first step, we calculate the general Ricci scalar expression for the MEMe Model. Tracing field equations (\ref{EFQMEMe}), one obtains
\begin{equation}\label{ricciexp}
R = -\kappa (-T_1+4T_2)=-\frac{4\kappa}{q}(|A|-1)\,,   
\end{equation}
where $|A|$ is given by (\ref{GCA-ExpAdet}).
From the conservation equation (\ref{ConsLaw}) it is possible to draw a relation for $\dot\rho$ that, in the case $w=1/3$, reads
\begin{equation}\label{rhodot}
{\dot \rho}\Big|_{w=1/3}=-4H\rho \, \frac{3-q\rho}{3-5q\rho}\,.
\end{equation}
By combining (\ref{ricciexp}) and (\ref{rhodot}), we can immediately derive the general expression for the time derivative of the Ricci scalar for the radiation domination phase: 
\begin{equation}\label{dtricci}
\dot{R}=\frac{64 \kappa q  \rho^2 (q \rho-3)^3}{27 (q \rho+1) (5 q \rho-3)}H\,.
\end{equation}
As one can see, this quantity depends only on $H$, $\rho$, and the model cutoff scale $q$; it is evidently different from zero provided that $\rho\neq3/q$. The Hubble flow $H$ can be deduced by Friedman equation (\ref{FriedEq}) and, for flat Universe  $k=0$, is given by 
\begin{equation}\label{HMEMe}
H^2=\frac{1}{9} \{27 \Lambda - \kappa q \rho^2 (q \rho+1)[q \rho (q \rho-8)+18]+27\kappa \rho\}\,.
\end{equation}
Finally, plugging (\ref{dtricci}) into the baryon asymmetry formula (\ref{asym1}), one obtains
\begin{equation}\label{etaMeme}
\eta =-\frac{80 \kappa q H \rho^2 (q \rho-3)^3}{9 \pi ^2 g M_*^2 {\mathcal T}_D (q \rho+1) (5 q \rho-3)},
\end{equation}
that represents the baryon asymmetry parameter for the MEMe model in the gravitational baryogenesis approach. 

To evaluate (\ref{etaMeme}), one must evaluate the energy density $\rho$ of matter in its rest frame. Since we are interested in the radiation domination era, it seems thermodynamically reasonable to set the energy density of the unique matter fluid as the energy density of relativistic particles provided by the Boltzmann equation \cite{kolb,Dodelson03}, 
\begin{equation}\label{rhorad}
\rho({\mathcal T})=\frac{\pi^2 g_*}{30}\, {\mathcal T}^4\,.
\end{equation}
We are now ready to use (\ref{etaMeme}) to put some new limits on the parameter $q$ that controls the matter coupling. To derive this constraint we set ${\mathcal T}={\mathcal T}_D$ in Eq.\eqref{rhorad} and take into account the measured baryon asymmetry \cite{AharonyShapira:2021ize}

\begin{equation}\label{upper}
\eta \lesssim \eta_{obs}\sim  10^{-10}\,.
\end{equation}
In order to get some quantitative results, we need to set the values of the cutoff mass and the decoupling temperature. To span a different range of conditions, following \cite{gravbar}, we chose two pairs of values. We initially consider ${\mathcal T}_D=10^{16}GeV$, $M_*=10^{18}GeV$ so that ${\mathcal T}_D$ is far from $M_*$ and, then, a second case in which we set ${\mathcal T}_D=10^{15}GeV$, $M_*=10^{15}GeV$, i.e., they have similar values. This last choice corresponds to an upper bound on these parameters, which is still compatible
with PPN and other low-energy phenomenological results.
On the other hand, since, at present, $q\rho\ll 1$ we have $\Lambda_{eff}\approx \Lambda$, and we can set $\Lambda \approx  10^{-84} GeV^2$, i.e., set it to the value of the observed cosmological constant\footnote{Planck observations provide a best fit value for density parameter $\Omega_\Lambda = 0.6889\pm 0.0056$ and for the Hubble constant $H_0 = 67.66\pm 0.42 (km/s)/Mpc = (2.1927664\pm 0.0136) \times 10^{-18} s^{-1}$ that allow to calculate the reported value of $\Lambda$ \cite{Planck:2018vyg,Abdalla:2022}.
However, one would be entitled to use such a value only if the analysis of the CMB fluctuations in this context is performed. For this reason, we rely on the estimation based on Supernovae type Ia in \cite{DES2024}, for which the corrections of the MEMe model can be deemed not relevant, given the bounds on $q$ in \eqref{GWqConstraint} and \eqref{GCAqConstraint}. In practice, given the approximations employed in the baryogenesis, the difference between the two values is irrelevant.
According to the other parameters approximation, we only use here two significant digits.}.
\begin{figure}[h!]
    \centering
    \includegraphics[scale=0.5]{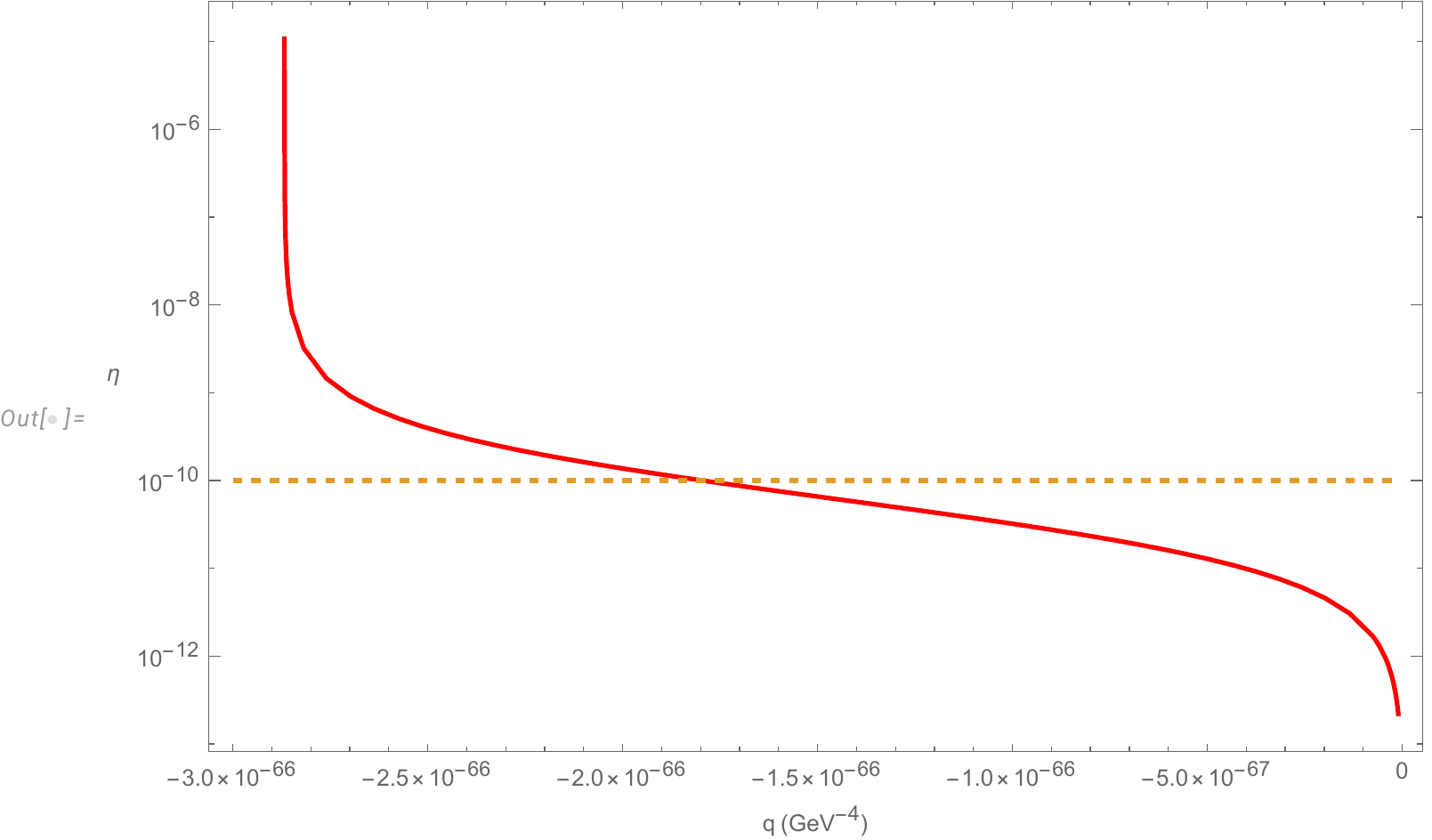}
    \caption{A logplot of $\eta$ vs. $q$ for ${\mathcal T}_D=10^{16}GeV$ and $M_*= 10^{18}GeV$. The cosmological constant is fixed to its observational value as provided by Planck.}
    \label{fig:eta_q_R}
\end{figure} 

\noindent From Fig.~\ref{fig:eta_q_R} it is evident that, with ${\mathcal T}_D=10^{16}GeV$, $M_*=10^{18}GeV$, in order to satisfy the condition \eqref{upper} we need to have negative $q$ and
\begin{equation}\label{betafinale}
|q| \lesssim 2\times10^{-66}GeV^{-4}\,.
\end{equation}
This result is both unexpected and tantalizing. In fact, the inverse of $|q|$, calculated in correspondence with the upper limit given in Eq.(\ref{betafinale}), determines an energy density that is close to (but lower than) the quantum vacuum energy density $\rho_{\Lambda}^{theory}\sim 10^{72} \div 10^{76}GeV^4$ \cite{Martin:2012bt}, \cite{Burgess:2013ara}). Such an outcome supports the interpretation of $q$ as a quantity related to the quantum vacuum for matter fields, and, in this perspective, it seems to confirm our initial interpretation of the MEMe model as a classical, coarse-grained representation of quantum field theory in curved spacetime. In such a picture, $\lambda=\kappa/q$ provides a natural regularization scale for the vacuum energy of quantum fields. The fact that beyond this energy scale, the Jordan frame of the theory loses meaning simply means that at those energies, a generalized coupling of the form chosen for the MEMe model is no longer sufficient to describe the quantum regime, even if the theory in the Einstein frame is still well defined. In addition, when $q\rho\approx 1$, we have  $\Lambda_{eff}\approx -\lambda>0$ and therefore the regularization scale for quantum fields is naturally tied to the cosmological constant term.

\begin{figure}[h!]
\centering
\begin{subfigure}{\textwidth}
  \centering
  \includegraphics[scale=0.27]{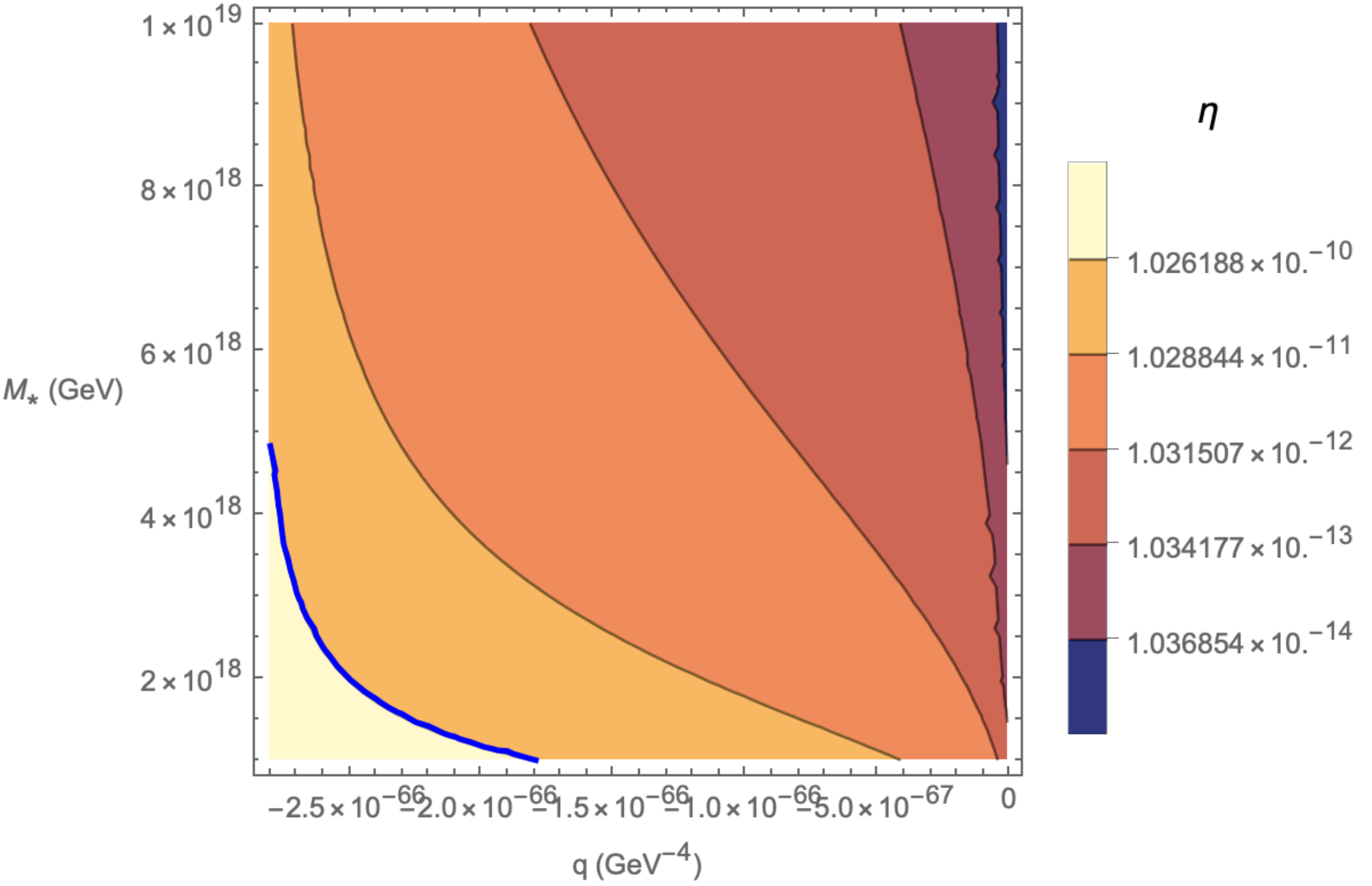}
    \caption{A contour plot of $\eta$ as a function of $q$ and $M_*$ with ${\mathcal T}_D=10^{16}GeV$, $g_*= 106$ and $\Lambda=4.3 \times 10^{-84}GeV^2$.}
\end{subfigure}%
\\
\vspace{1cm}
\begin{subfigure}{\textwidth}
  \centering
  \includegraphics[scale=0.27]{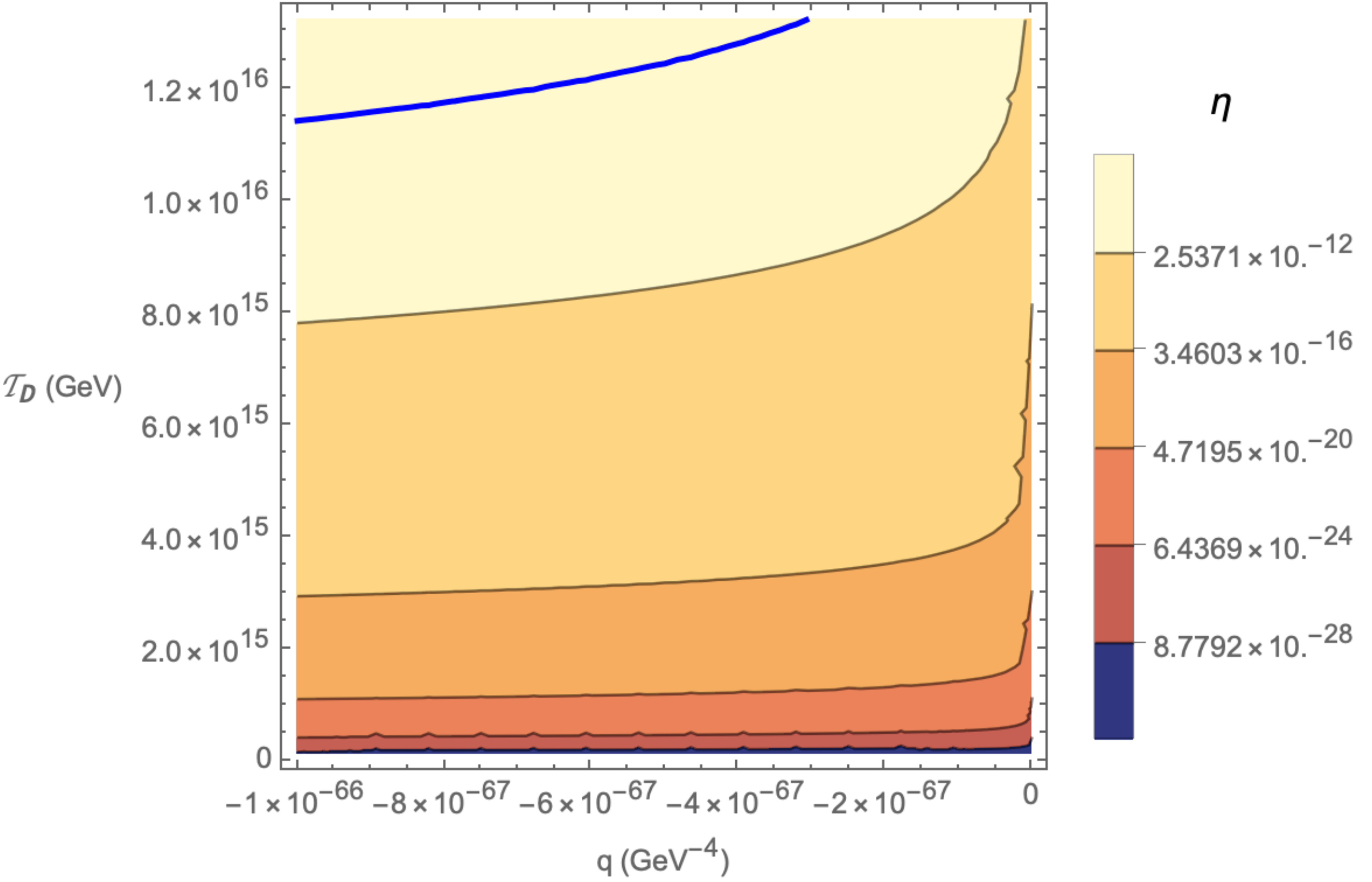}
    \caption{A contour plot of $\eta$ as a function of $q$ and $\mathcal{T}_D$ with $M_*=10^{18}GeV$, $g_*= 106$ and $\Lambda=4.3 \times 10^{-84}GeV^2$.}
\end{subfigure}
\caption{The expected baryon asymmetry in the MEMe model. The blue curve evidences the contour line for $\eta=10^{-10}$.}
\label{eta_q_T_M_R}
\end{figure}

\begin{figure}[h!]
\centering
\includegraphics[scale=0.27]{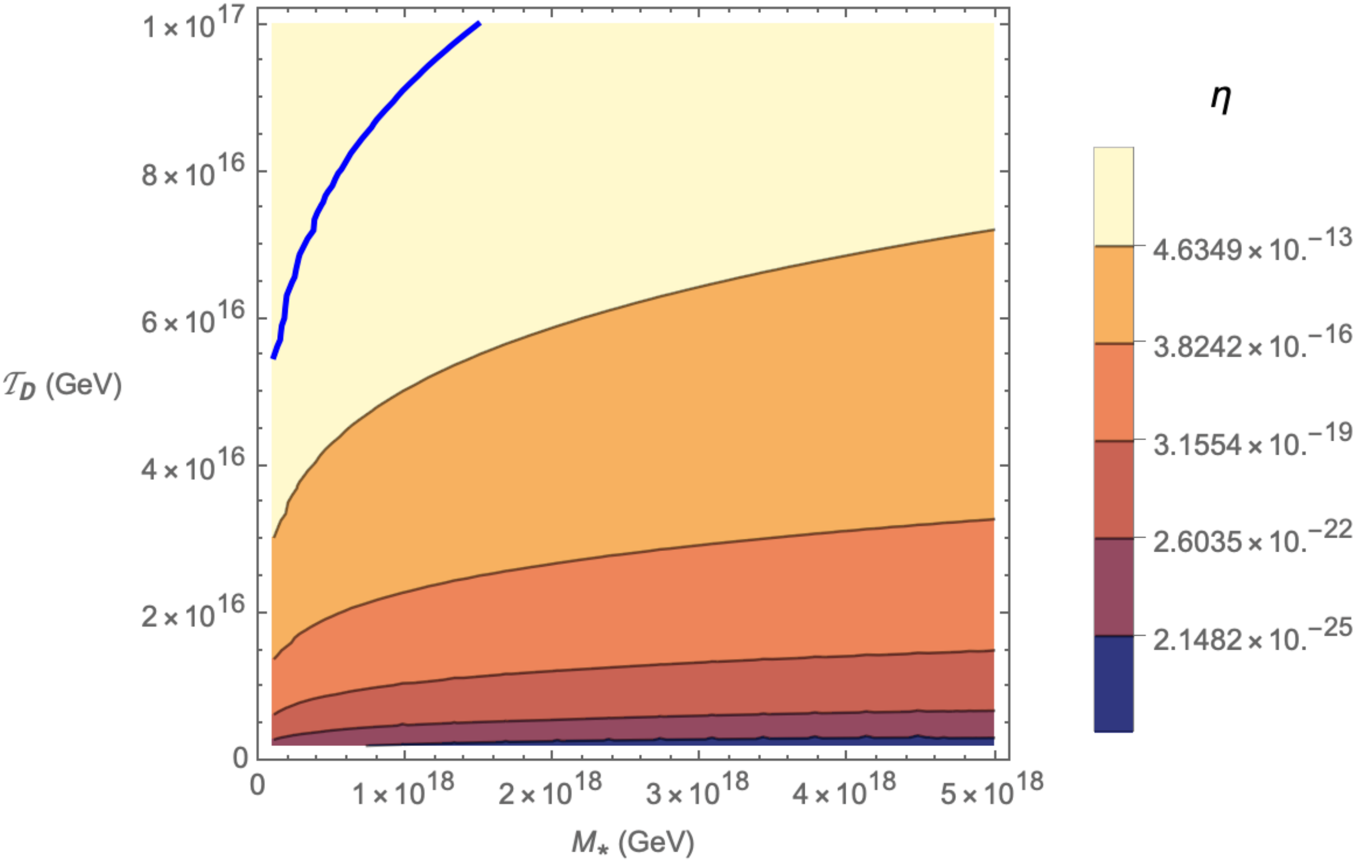}
\caption{The baryon asymmetry amount $\eta=\eta({\mathcal T}_D\,,M_*)$ by assuming $q=-10^{-74}GeV^{-4}$. 
As before, $g_*= 106$ and $\Lambda=4.3 \times 10^{-84}GeV^2$. The blue curve evidences the contour line for $\eta=10^{-10}$.}
\label{fig:qcrit_R}
\end{figure}

To further understand the baryogenesis constraints, we have then studied (\ref{etaMeme}) varying the decoupling temperature and the cutoff mass scale. In Fig.~\ref{eta_q_T_M_R} and \ref{fig:qcrit_R}, we display some contour plots that show the behavior of $\eta$ as a function of model parameters. More specifically, in Fig.~\ref{eta_q_T_M_R} we show  $\eta=\eta(q,\,M_*)$ ({\it upper panel}) and $\eta=\eta(q,\,{\mathcal T}_D)$ ({\it lower panel}) respectively. From these plots, one can infer that, if $\mathcal{T}_D=10^{16}GeV$, a suitable baryon asymmetry is achieved when $M_*$ spans less than one order of magnitude around $10^{18}GeV$. On the other hand, if the cutoff mass is set to $M_*=10^{18}GeV$, we can have $\eta=10^{-10}$ when $\mathcal{T}_D$ slightly differs from $10^{16}GeV$, provided a suitable value for $q$.

In Fig.~\ref{fig:qcrit_R}, we have set $q=-10^{-74}GeV^{-4}$ and studied the variation of $\eta$ varying both ${\mathcal T}_D$ and $M_*$. Our choice of $q$ is a value that, as seen above, can be considered in agreement with the inverse of the theoretical zero-point energy of quantum field theory. The plot suggests that once a very small value of $q$ is considered, a matter-antimatter imbalance of order $10^{-10}$ is achieved with decoupling temperatures higher than $5\cdot10^{16}GeV$ and cutoff masses about $10^{18}GeV$.
In general, all these plots evidence that the amount of baryon asymmetry can agree with data in a significant region of the parameter space. 

As it is often found in literature, to get the model prediction in a lower energy regime limit, we have also studied the model when  ${\mathcal T}_D=10^{15}GeV$ and $M_*=10^{15}GeV$. Below, we provide the different graphics that highlight the results in this case. 
\begin{figure}[h!]
    \centering
    \includegraphics[scale=0.3]{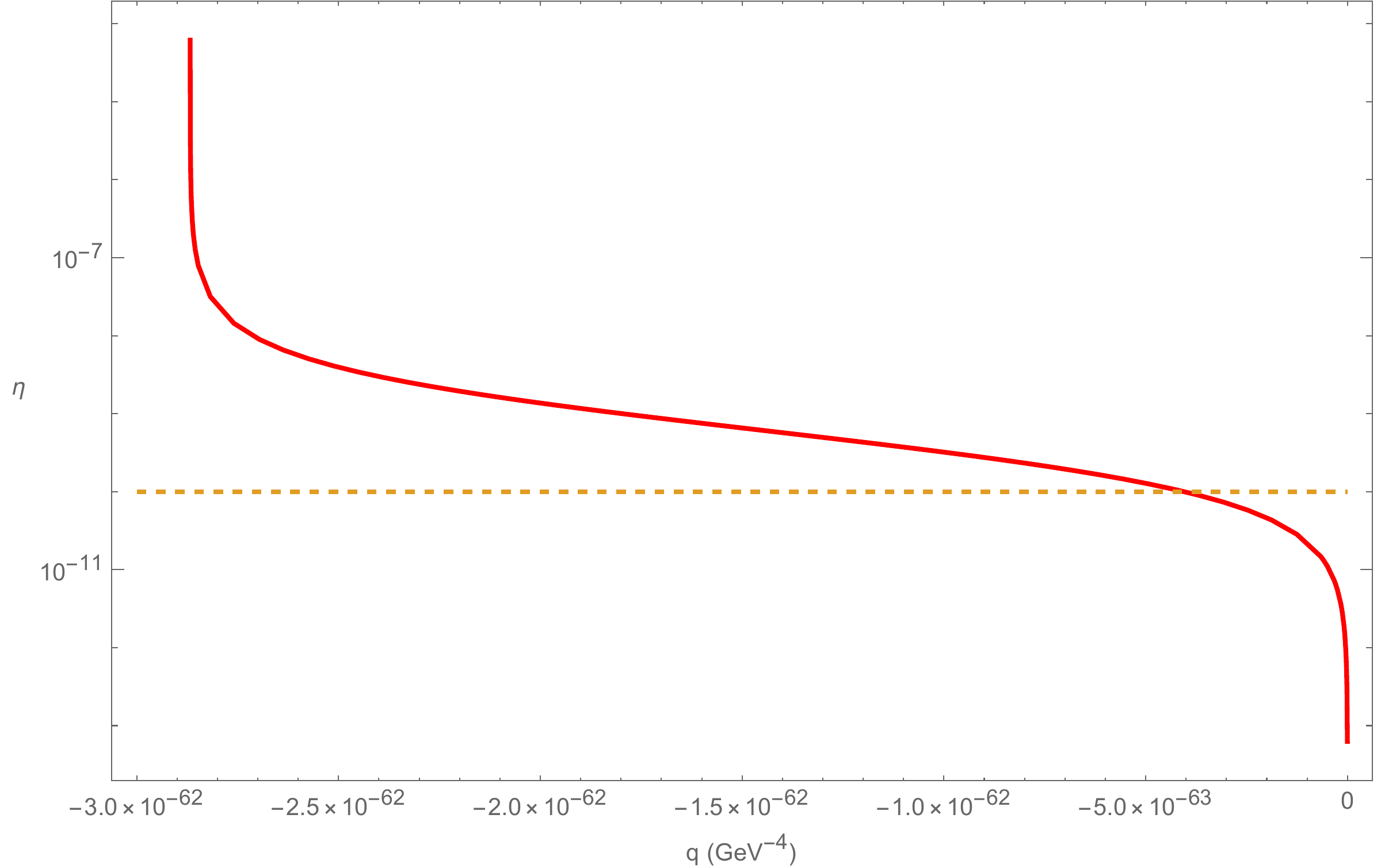}
    \caption{A logplot of $\eta$ vs. $q$ for ${\mathcal T}_D=10^{15}GeV$ and $M_*= 10^{15}GeV$. The cosmological constant is fixed to its observational value as provided by Planck.}
    \label{fig:eta_q}
\end{figure}
As one can see, the observed baryon asymmetry prediction is obtained at lower magnitudes of the model parameter $q$:
\begin{equation}\label{betafinale2}
|q| \lesssim 4\times10^{-63}GeV^{-4}\,.
\end{equation}
Hence, lowering the values of the mass cutoff scale and the decoupling temperature, the observed baryon asymmetry is obtained for a value of the parameter $q$ that is three or four orders of magnitude smaller than before.
\begin{figure}
\centering
\begin{subfigure}{\textwidth}
  \centering
  \includegraphics[scale=0.23]{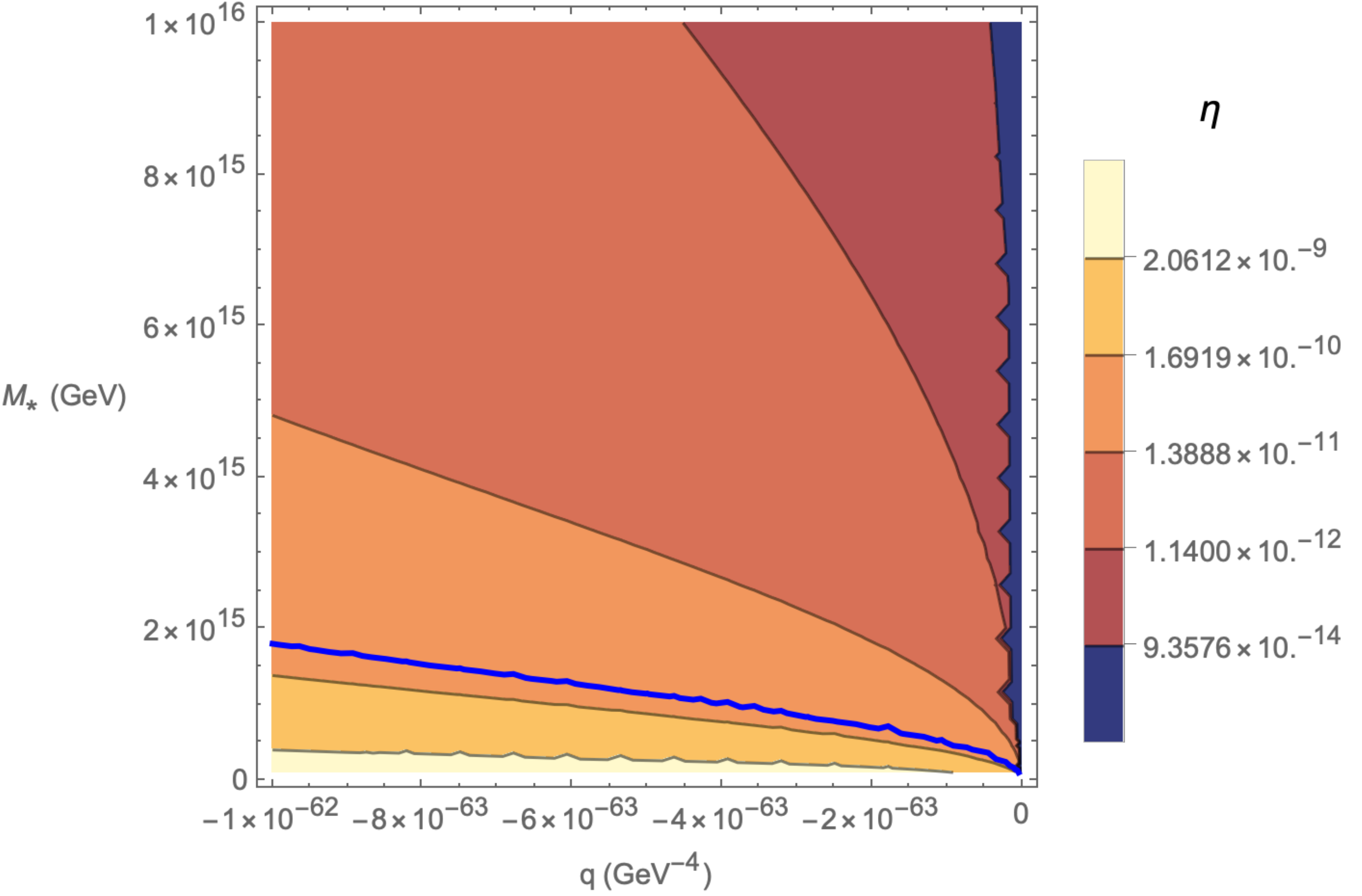}
    \caption{A contour plot of $\eta$ as a function of $q$ and $M_*$ with ${\mathcal T}_D=10^{15}GeV$, $g_*= 106$ and $\Lambda=4.3 \times 10^{-84}GeV^2$.}
\end{subfigure}%
\\
\vspace{1cm}
\begin{subfigure}{\textwidth}
  \centering
  \includegraphics[scale=0.23]{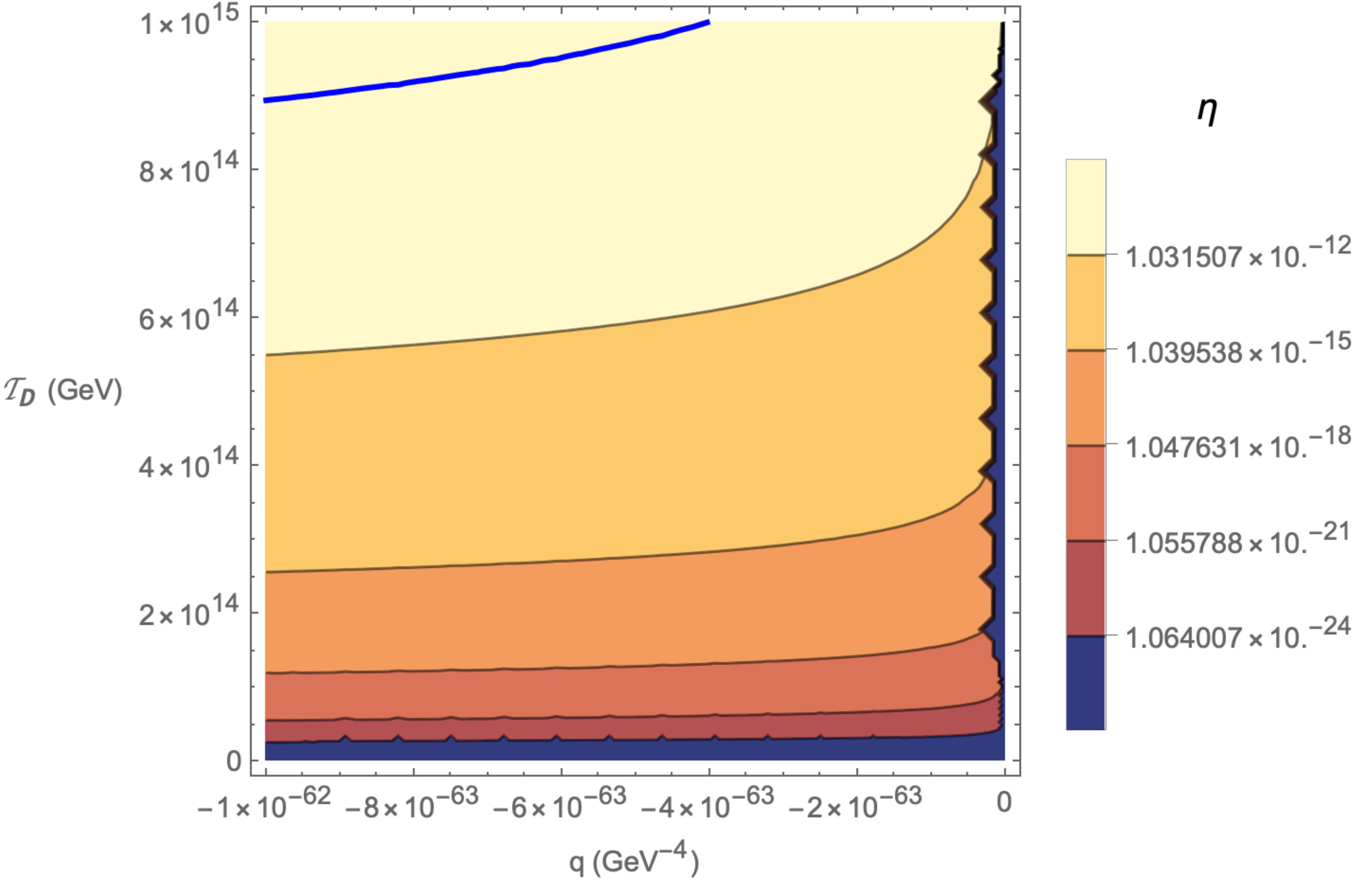}
    \caption{A contour plot of $\eta$ as a function of $q$ and $\mathcal{T}_D$ with $M_*=10^{15}GeV$, $g_*= 106$ and $\Lambda=4.3 \times 10^{-84}GeV^2$.}
\end{subfigure}
\caption{The expected baryon asymmetry in the MEMe model. The blue curve evidences the contour line for $\eta=10^{-10}$.}
\label{eta_q_T_M}
\end{figure}

\begin{figure}[h!]
\centering
\includegraphics[scale=0.25]{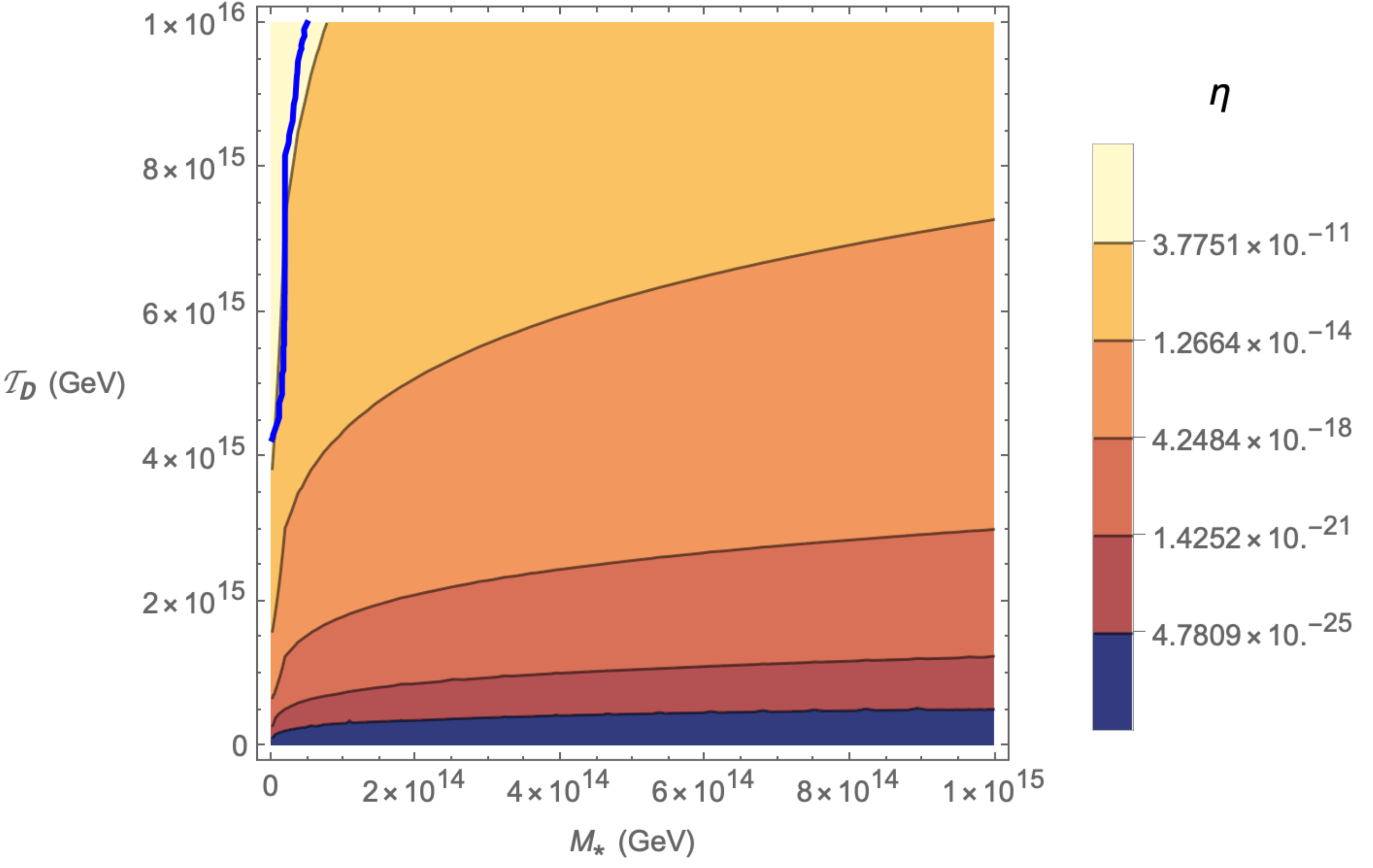}
\caption{The baryon asymmetry amount $\eta=\eta({\mathcal T}_D\,,M_*)$ by assuming $q=-10^{-74}GeV^{-4}$. 
As before, $g_*= 106$ and $\Lambda=4.3 \times 10^{-84}GeV^2$. The blue curve evidences the contour line for $\eta=10^{-10}$.}
\label{fig:qcrit}
\end{figure}

\subsection{The spatially curved case}

In this section, we extend the analysis to the case of a non-flat-FRW cosmology with $k\neq 0$. Since, as we have seen, in the Einstein frame of the MEMe model the usual cosmological eras cannot be separated as in the classical GR cosmologies, it is necessary to investigate also the role of spatial curvature on baryogenesis. Following the same reasoning as in the previous section, one starts from the general form of field equations (\ref{FriedEq}) and (\ref{RayEq}) and writes the relations for $\dot{R}$ and $\eta$ that generalize Eqs. (\ref{dtricci}) and (\ref{etaMeme}). The Hubble parameter is now given by:
\begin{equation}\label{hubblek}
    H^{2}=\frac{1}{9} \{27 \Lambda - \kappa q \rho^2 (q \rho+1)[q \rho (q \rho-8)+18]+27\kappa \rho\}-\frac{k}{S^2}\ ,
\end{equation}
and should be substituted into Eq.\eqref{etaMeme}.
Unlike the previous case, the scale factor $S$ explicitly appears in the expression for $\eta$. As a consequence, we need to connect $S$ to known quantities in order to estimate $\eta$. Actually, GR provides a well-known relation between the scale factor and the temperature of the cosmic fluid derived by entropy conservation ($S\propto \mathcal{T}^{-1}$). Therefore, one could be tempted to use it to solve the problem. However, this would result in a mistake. The classical relations that connect the scale factor to $\mathcal T$ are derived under the assumption that the observers are comoving with matter, whereas we are not dealing with our cosmological model from that perspective. 

The correct way to proceed is first to use the relation between the scale factor $S$ and the Jordan frame scale factor $\mathcal{S}$ in \eqref{SEtoSJ}, which is related to observers comoving with matter, and then to use the conservation of entropy to connect $\mathcal{S}$ to the temperature.

At this point, entropy conservation during the radiation era provides the appropriate connection between the comoving scale factor and the temperature:
\begin{equation}
\mathcal{S}(t)=\mathcal{S}_{rad}\frac{{\mathcal T}_{rad}}{{\mathcal T}}\,,
\end{equation}
here $\mathcal{S}_{rad}$ and ${\mathcal T}_{rad}$ represent the value of the comoving scale factor and of the temperature at radiation domination.  

We can obtain a bound on these two quantities by considering the matter radiation equality in the context of the MEMe model. In particular, assuming, as we have done, that matter is a perfect fluid, \eqref{ConsJ} implies that
\begin{equation}\label{rhoSJ}
\rho= \frac{\rho_0}{ \mathcal{S}^{3(w+1)}}
\end{equation}
where $\rho_0$ is an integration constant that represents the value of the energy density at a given time.  As a consequence, the value of the Jordan scale factor at matter radiation equality can be obtained by setting $\rho_{d}=\rho_{r}$ when  $\rho_{d}$ is given by \eqref{rhoSJ} with $w=0$ and $\rho_{r}$ is given by \eqref{rhoSJ}  for $w=1/3$. We obtain then
\begin{equation}
\mathcal{S}^{eq}=\frac{\rho_{0,r}}{\rho_{0,d}}
\end{equation}
In GR, the constants $\rho_{0,d}$ and $\rho_{0,r}$ are normally set at the values of the radiation and dust density observed today, as we assume the scale factor to be normalized at its current value. In the case of the MEMe model (assuming the same normalization), these values are essentially the same. In fact, as mentioned before, there is no difference between the cosmic energy density measured from Earth and $\rho$ as they are both defined in the Jordan frame. Hence, we can write
\begin{equation}
\mathcal{S}^{eq}\approx\frac{\bar{\rho}_{0,r}}{\bar{\rho}_{0,d}}
\end{equation}
where the bar indicates the quantities observed from Earth, and we can finally set that $\mathcal{S}_{eq}$ and $\mathcal{T}_{eq}$  are approximated by their GR counterparts (see e.g. \cite{AharonyShapira:2021ize} for an estimate of their values). Using these results we can establish an higher bound for $\mathcal{S}_{rad}$ and a lower bound of ${\mathcal T}_{rad}$  as
\begin{equation}
\mathcal{S}_{rad}<\mathcal{S}_{eq}=1/1+z_{eq} \quad \mbox{with} \quad z_{eq}\approx3400
\end{equation}
and, by using the Boltzmann equation,
\begin{equation}
{\mathcal T}_{rad}\gtrsim3.4\times10^4K\approx  10^{-9}GeV.
\end{equation}
The above reasoning allows us to write $H$ in terms of ${\mathcal T}$ and evaluate $\eta$ in the non-spatially flat case. 
As in the spatially flat case, we explore the model prediction capability considering two different couples of values for $M_*$ and $\mathcal T_D$. At first, we set $M_*=10^{18}GeV$ and $\mathcal T_D = 10^{16}GeV$. 
\begin{figure}[h!]
%\begin{subfigure}{.5\textwidth}
  \centering
  \includegraphics[scale=0.21]{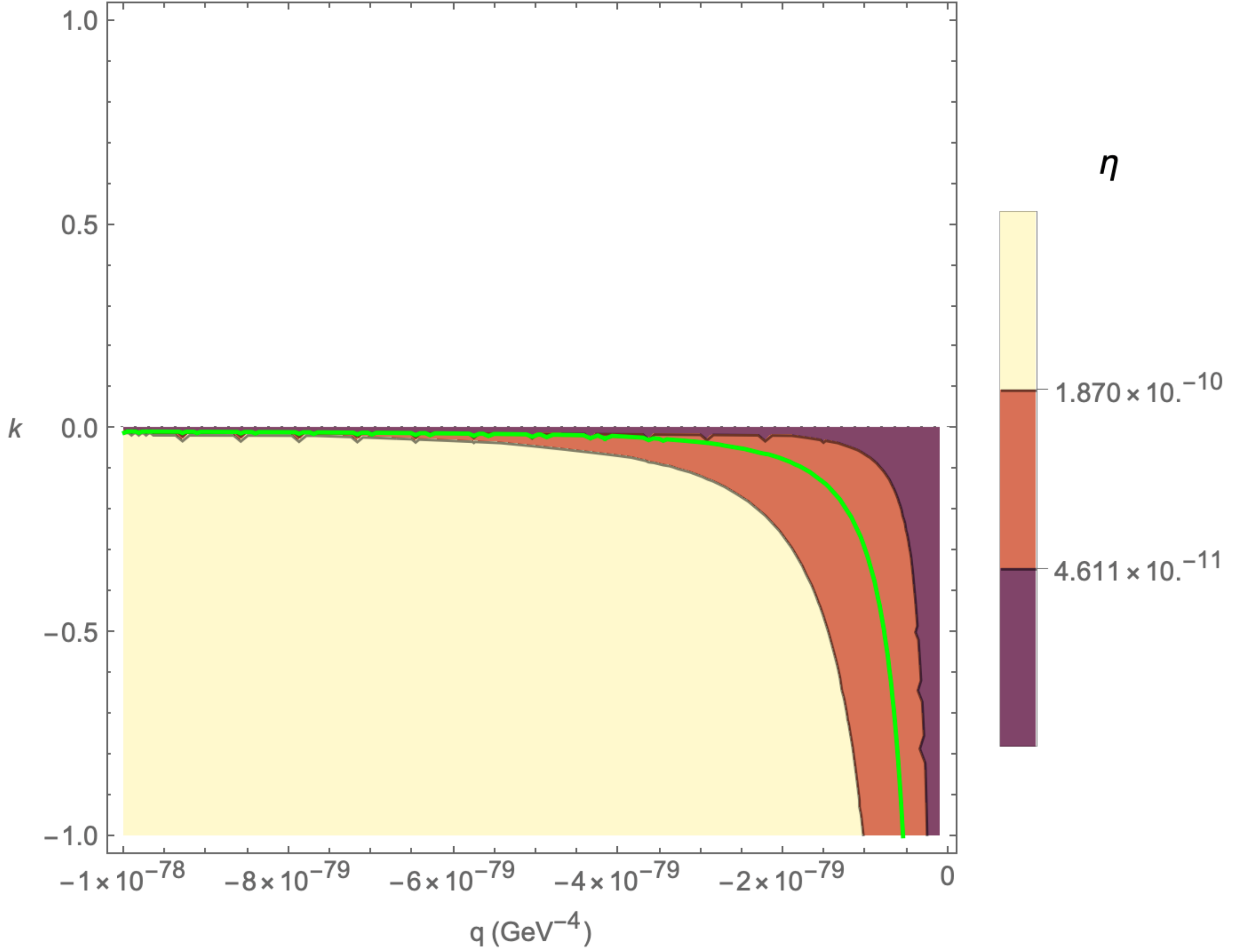}
%\end{subfigure}
\caption{A contour plot displaying the behavior of $\eta$ as a function of $q$ and $k$. Other parameters are ${\mathcal T}_D=10^{16}GeV$, $M_*=10^{18}GeV$ $g_*= 106$ and $\Lambda=4.3\times 10^{-84}GeV^2$. The green curve evidences the contour line for $\eta=10^{-10}$. 
It is clear that the data favor a flat spatial geometry. It is possible to have an open universe (with $-1<k<-0$) only for $-10^{-80}GeV^{-4}\lesssim q < 0$ .}
\label{eta_q_k_R}
\end{figure}
Fig.~\ref{eta_q_k_R} highlights that, for most of the values of $q$, only flat ($k=0$) cosmologies are compatible with the limits on baryogenesis. Comparison of $\eta=\eta(k)$ with data also excludes closed ($k=1$) universes.
\begin{figure}[h!]
\centering
\includegraphics[scale=0.4]{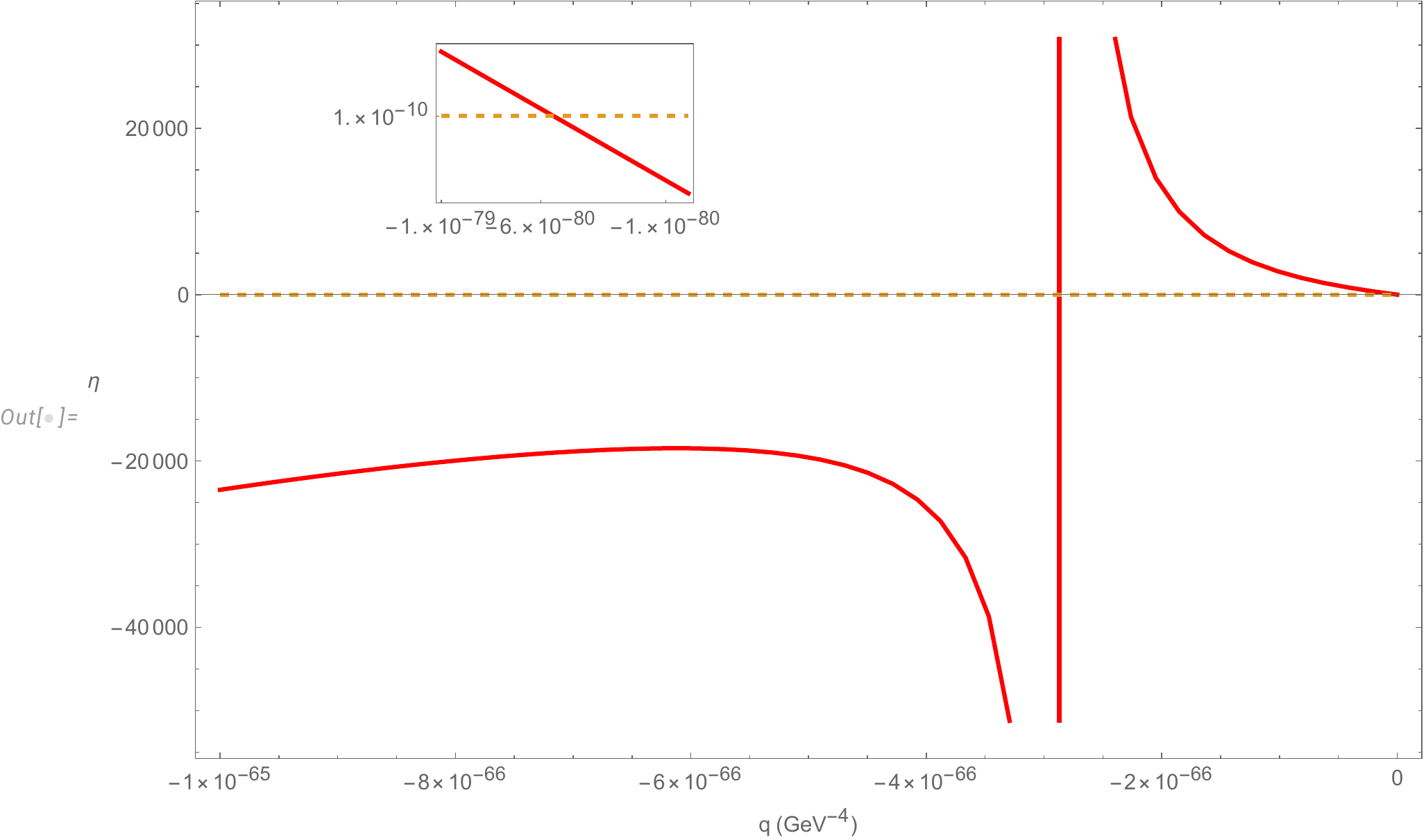}
\caption{Baryon asymmetry for a open universe ($k=-1$) vs. $q$ with $M_*=10^{18}GeV$ and $\mathcal{T}_D=10^{16}GeV$. The orange dashed line gives  $\eta=10^{-10}$. In the small panel, the value of $q$ for which $\eta$ is $10^{-10}$.}
\label{fig:eta_q_kopen_R}
\end{figure}
Fig.~\ref{fig:eta_q_kopen_R} shows the behavior of baryon asymmetry for an open ($k=-1$) universe. To have $\eta \lesssim 10^{-10}$, very small values of the parameter $q$ are required. More specifically, an open universe is consistent with the baryogenesis bound only if $q$ is negative and its modulus is
\begin{equation}
|q|\lesssim 5\times10^{-80} GeV^{-4}
\end{equation}
a value that pushes the energy density related to $q$ to take values beyond the zero point energy density of quantum field theory.
\begin{figure}[h!]
\centering
\begin{subfigure}{\textwidth}
  \centering
  \includegraphics[scale=0.23]{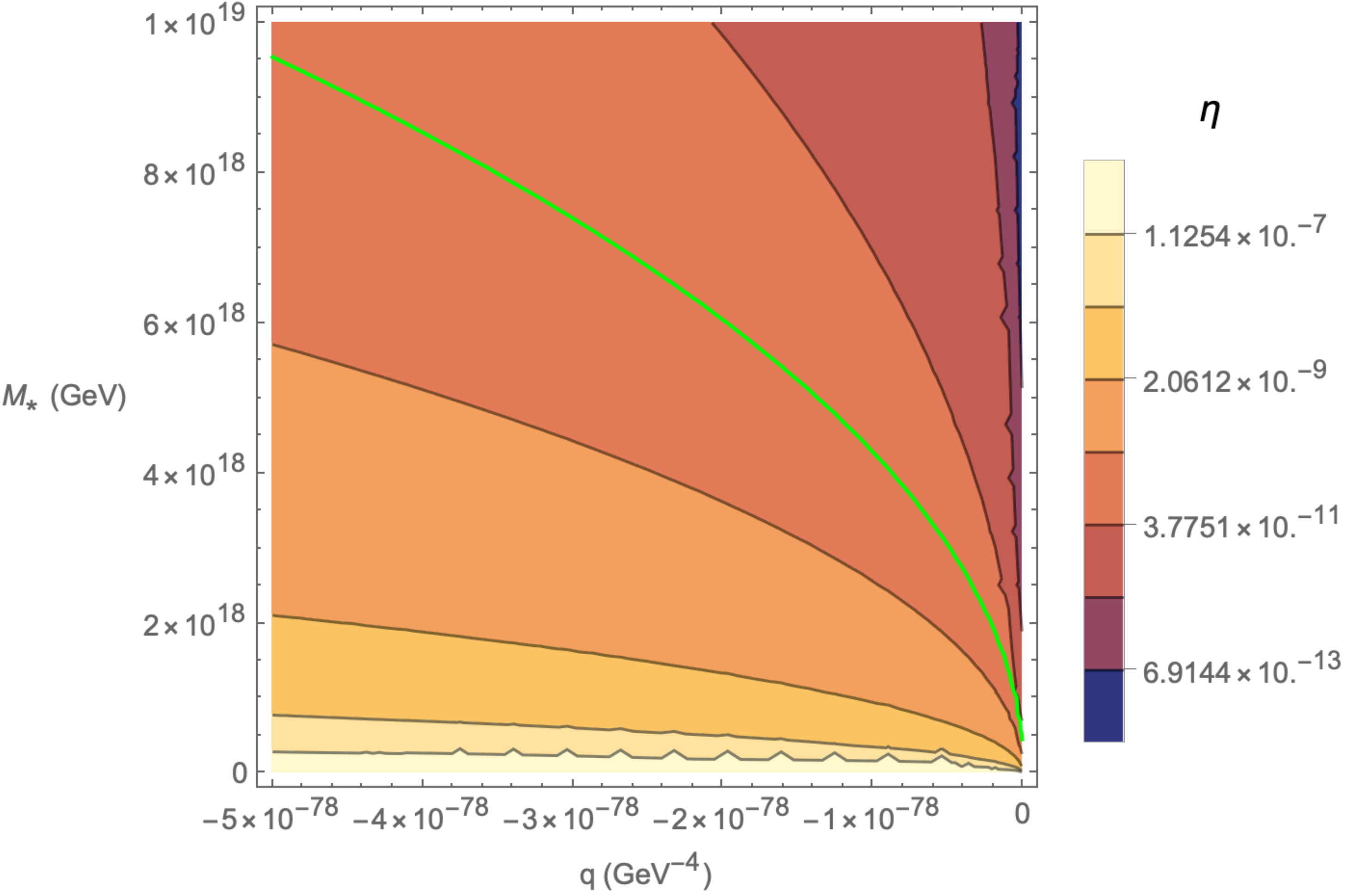}
    \caption{The baryon asymmetry $\eta$ as a function of $q$ and $M_*$ with ${\mathcal T}_D=10^{16}GeV$.}
\end{subfigure}%
\\
\vspace{1cm}
\begin{subfigure}{\textwidth}
  \centering
  \includegraphics[scale=0.23]{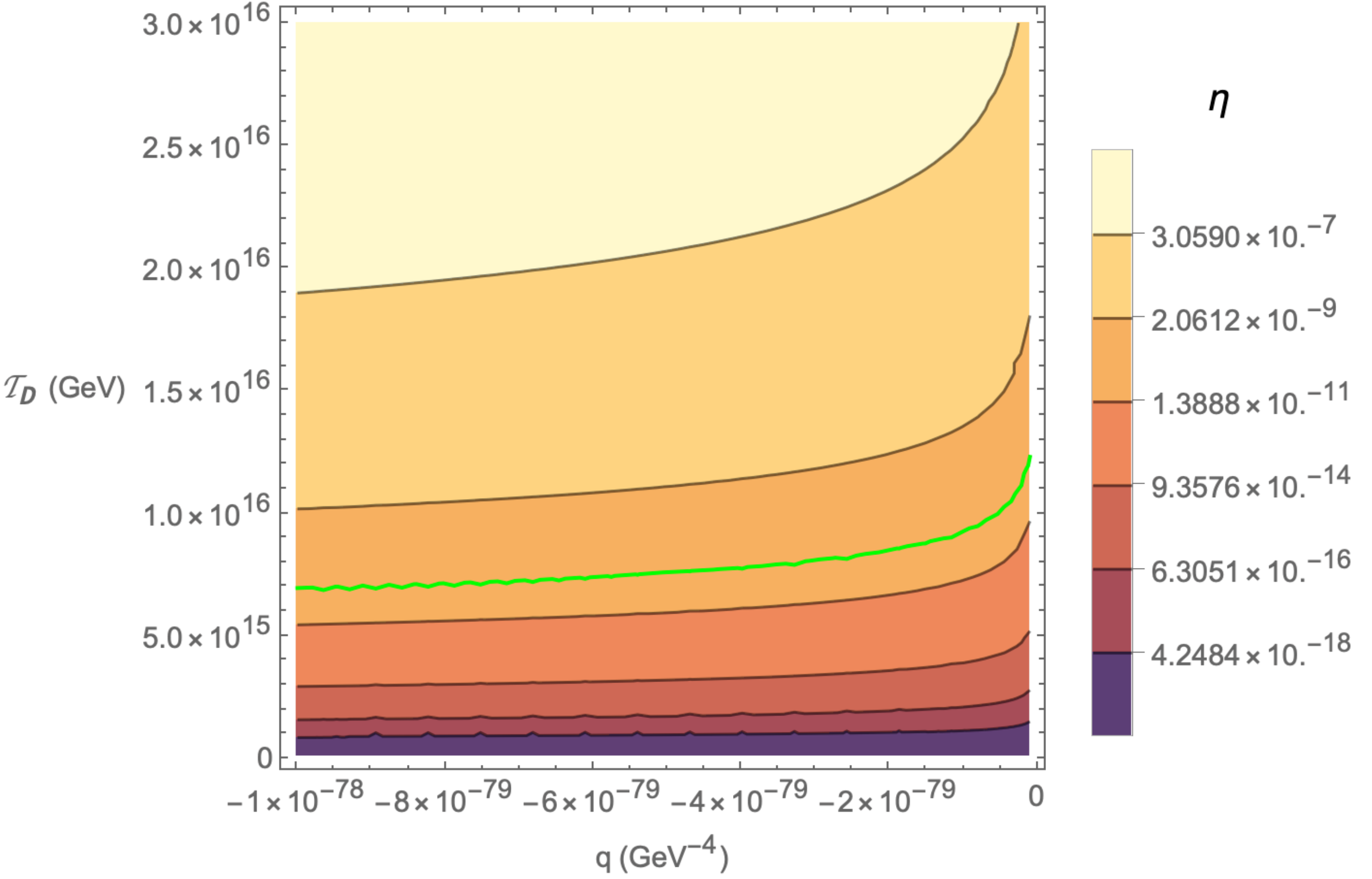}
\caption{The baryon asymmetry $\eta$ as a function of $q$ and ${\mathcal T}_D$ with $M_*=10^{18}GeV$.}
\end{subfigure}
\caption{Baryon asymmetry model for an open universe ($k=-1$). Other parameters are $g_*= 106$ and $\Lambda=4.3 \times 10^{-84}GeV^2$. The green curve evidences the contour line for $\eta=10^{-10}$.}
\label{eta_kopen_T_M_R}
\end{figure}
In Fig.~\ref{eta_kopen_T_M_R}, we show the MEMe model baryon asymmetry prediction for $k=-1$ varying, respectively, $M_*$ and ${\mathcal T}_D$ with the model scale $q$. We can see that if one considers $\mathcal{T}_D=10^{16}GeV$, cutoff masses $M_*$ lower than $10^{19}GeV$ return $q$ to very tiny values. 
A similar behavior is obtained when one sets $M_*=10^{18}GeV$ and the decoupling temperature is assumed to be higher than $10^{16}GeV$.

In the same manner as the spatially flat case, the whole analysis is performed considering also the couple of values $M_*=10^{15}GeV$ and $\mathcal{T}_D=10^{15}GeV$. Again, it is possible to draw the observed baryon asymmetry provided that the parameter $q$ assumes the values
\begin{equation}
    |q|\lesssim 6\times 10^{-78}GeV^{-4}.
\end{equation}
Spatially closed universes are also discarded in this case. Moreover, we find that, as a general trend, lowering the energy scale requires higher values of $q$.

\begin{figure}
%\begin{subfigure}{.5\textwidth}
  \centering
  \includegraphics[scale=0.21]{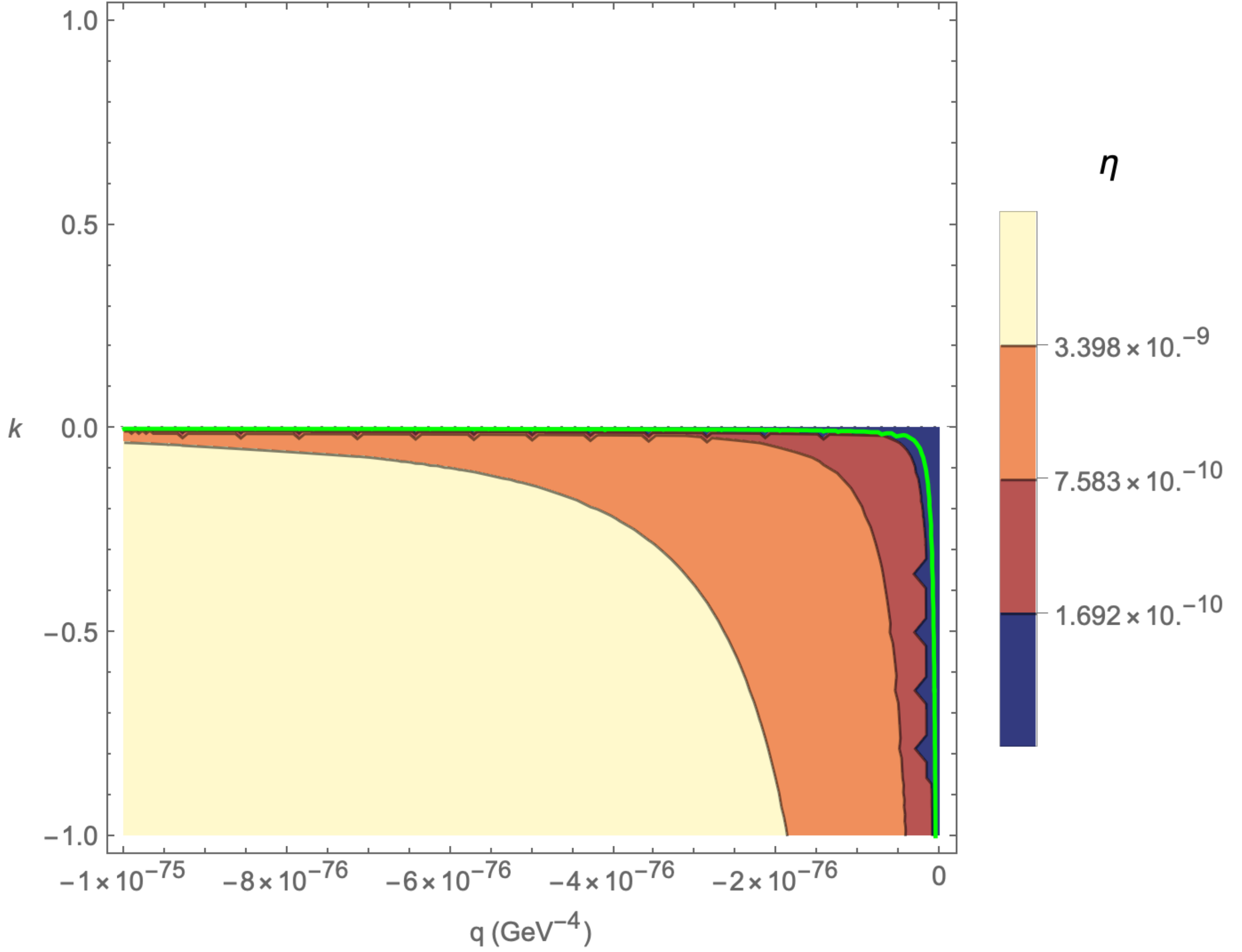}
%\end{subfigure}
\caption{A contour plot displaying the behavior of $\eta$ as a function of $q$ and $k$. Other parameters are ${\mathcal T}_D=10^{15}GeV$, $M_*=10^{15}GeV$ $g_*= 106$ and $\Lambda=4.3\times 10^{-84}GeV^2$. The green curve evidences the contour line for $\eta=10^{-10}$. 
It is clear that the data favor a flat spatial geometry. It is possible to have an open universe (with $-1<k<-0$) only for $  -10^{-78}GeV^{-4} \lesssim q < 0$}.
\label{eta_q_k}
\end{figure}
\begin{figure}[h!]
\centering
\includegraphics[scale=0.4]{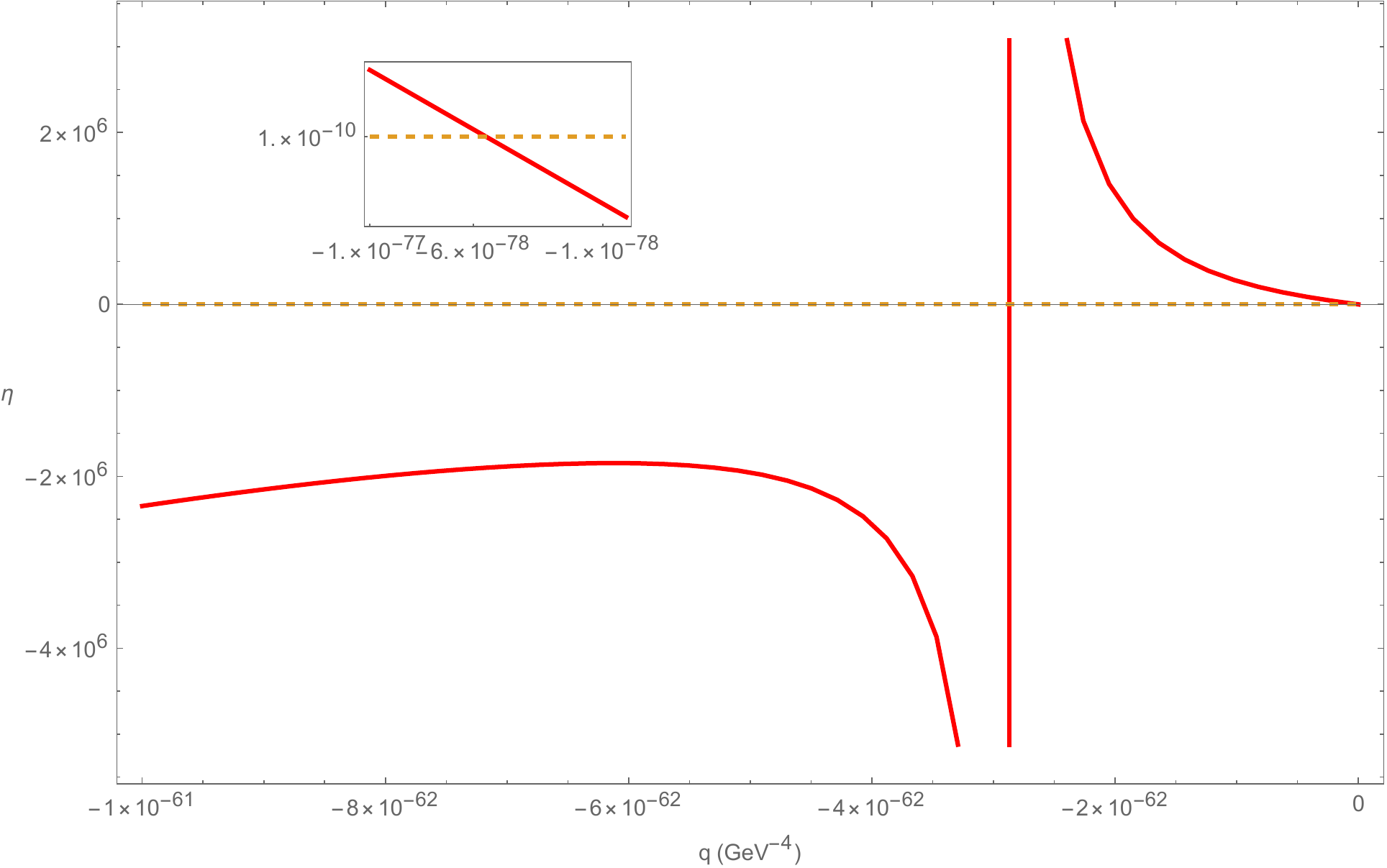}
\caption{Baryon asymmetry for a open universe ($k=-1$) vs. $q$, in this case $M_*=10^{15}GeV$ and $\mathcal{T}_D=10^{15}GeV$. Again, the orange dashed line gives  $\eta=10^{-10}$. In the small panel, the value of $q$ for which $\eta$ is $10^{-10}$.}
\label{fig:eta_q_kopen}
\end{figure}

\begin{figure}
\centering
\begin{subfigure}{\textwidth}
  \centering
  \includegraphics[scale=0.23]{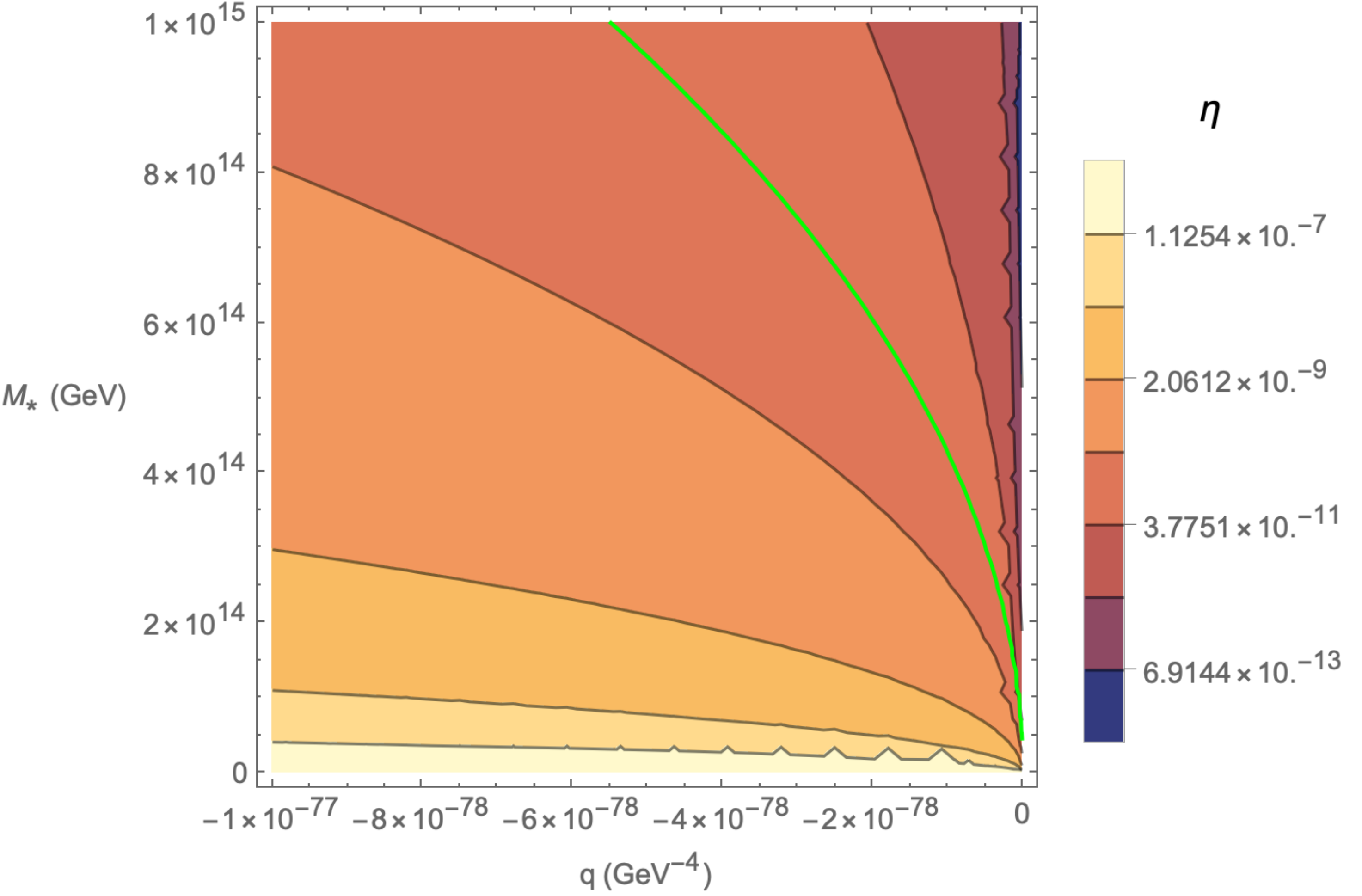}
    \caption{The baryon asymmetry $\eta$ as a function of $q$ and $M_*$ with ${\mathcal T}_D=10^{15}GeV$.}
\end{subfigure}%
\\
\vspace{1cm}
\begin{subfigure}{\textwidth}
  \centering
  \includegraphics[scale=0.23]{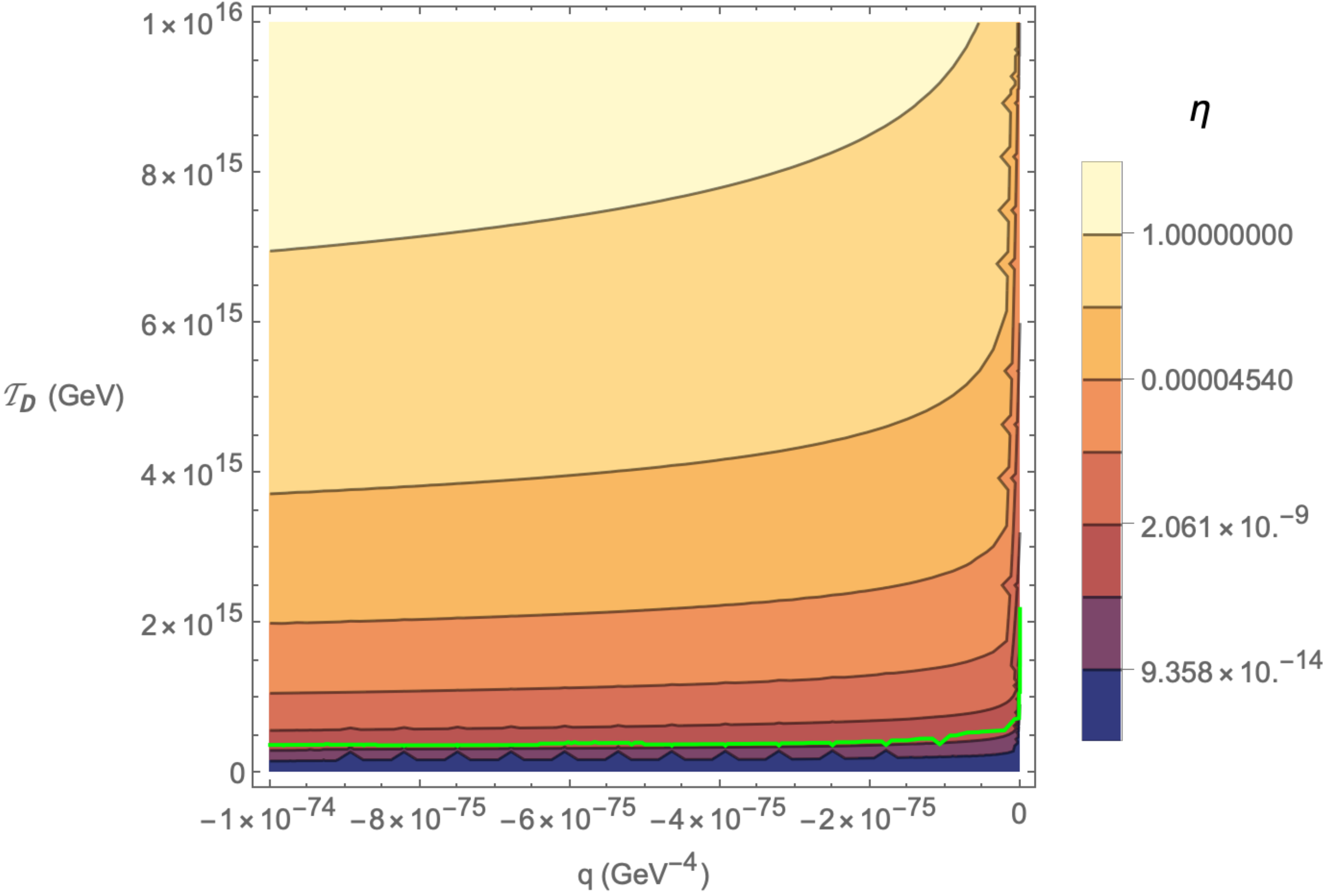}
\caption{The baryon asymmetry $\eta$ as a function of $q$ and ${\mathcal T}_D$ with $M_*=10^{15}GeV$.}
\end{subfigure}
\caption{Baryon asymmetry for an open universe ($k=-1$). Other parameters are $g_*= 106$ and $\Lambda=4.3 \times 10^{-84}GeV^2$. The green curve evidences the contour line for $\eta=10^{-10}$.}
\label{eta_kopen_T_M}
\end{figure}

\section{Conclusions}

In this paper, we have considered the early time cosmology of a new class of GR modifications, dubbed ``generalized coupling theories,'' and more specifically, the so-called Minimal Exponential Measure Model (MEMe). 

Generalized coupling theories are characterized by a nonlinear relation between the Einstein tensor and the stress-energy tensor. In the case of the MEMe model, this relation has the advantage of depending on just one parameter, $q$. In some previous works, this quantity has been constrained by employing Gravitational waves and Parametrized-Post-Newtonian phenomenology. To check the model's reliability at all scales, we studied the MEMe model's behavior in the very early universe. More specifically, since the MEMe model provides a non-vanishing Ricci scalar in the radiation domination era, it is possible to implement the gravitational baryogenesis mechanism within such a scheme. 

The gravitational baryogenesis approach assumes that the matter current is directly coupled to scalar curvature, and this non-minimal coupling is responsible for ``propagating" in time through thermal equilibrium the baryon asymmetry frozen at the decoupling temperature.  In deriving the actual form of the gravitational baryogenesis term in the context of the MEMe model,  we needed to consider that generalized coupling theories can be regarded as bimetric theories. One metric,  $g$, describes purely gravitational degrees of freedom, while the other metric, $\mathfrak{g}$, represents a geometry constructed on the basis of matter particle motions. This fact implies that gravitational phenomenology can be interpreted in two different (but equivalent) ways. This aspect is crucial in characterizing the gravitational baryogenesis parameter, as it raises the problem of the choice of the Ricci scalar appearing in such a term.  
We have argued that because of the mechanism underlying gravitational baryogenesis,
the Ricci scalar should be the one associated with $g$. 

Once we had chosen the theoretical framework, we deduced the constraints on $q$ resulting from the observed baryon asymmetry. Since the Einstein frame cosmological equations are not linear in the energy density, the classical division of cosmic history into different eras might not apply. Consequently, terms like the spatial curvature and the cosmological constant cannot be automatically neglected in the calculations on gravitational baryogenesis. However, as the spatially curved case requires some additional considerations, we decided to present it separately to ease the reading of the manuscript. 

Our analysis yields a very stringent constraint on the modulus of the parameter $q$. Considering a decoupling temperature $\mathcal T_D=10^{16}GeV$ and a mass scale for the effective theory of $M_*=10^{18}GeV$, we have obtained, in the spatially flat case, a negative $q$ with magnitude $|q|\lesssim 2\times 10^{-66}GeV^{-4}$. 
On the other hand, we found that only an open universe is compatible with baryogenesis in the limit $|q|\lesssim 5\times 10^{-80}GeV^{-4}$, which is in its absolute value a much tighter constraint. 

The bounds of $q$ associated with the choice $\mathcal T_D=10^{15}GeV$ a $M_*=10^{15}GeV$ are a few orders of magnitude different from the above. Specifically, we have $|q|\lesssim 4\times 10^{-63}GeV^{-4}$ in the flat case and $|q|\lesssim 6\times 10^{-78}GeV^{-4}$ in the (negatively) curved case. These results are robust in the sense that even working close to the cutoff values of the theory, the obtained value for $q$ remains coherent with the ones calculated in the case in which $\mathcal T_D$ is far from $M_*$.
 
In light of such outcomes, some interesting aspects of the MEMe model naturally come into view. First, as the energy densities typical of the baryogenesis processes, $\rho\approx 10^{60} GeV^{4}$, the bound on $q$ we have found indicates that  $q\rho\ll 1$ and, consequently,  $\mathfrak {g}$ and $g$ are very close to each other. In turn, this implies that $R(g)\approx R(\mathfrak {g})$, i.e., the error in using the ``wrong Ricci scalar'' in the gravitational baryogenesis term turns out to be very small with respect to the uncertainties in other quantities involved in the calculations. Secondly, the value of the energy density cutoff scale of the theory $\lambda/\kappa =1/q $ that is congruent with the baryogenesis constraint is quite reconcilable with the vacuum energy density of physical fields $\rho_{vac}^{theory}\sim 10^{72} \div 10^{76}GeV^4$. This fact suggests that the scale $q$ could indeed be associated with the onset of quantum effects in classical matter. The introduction of spatial curvature shifts the value of $q$ towards very tiny values that, in terms of the associated energy density, are again compatible with $\rho_{vac}^{theory}$ and in some sense beyond this value. Such an outcome suggests the fascinating scenario that, in this scheme, spatial curvature and the vacuum energy of the matter fields are intimately related.

These results, although very interesting, should be weighed with care. Gravitational baryogenesis is a fascinating mechanism to generate baryon asymmetry. Nevertheless, its origin depends on the feature of an underlying theory of gravity, whose details still need to be fully verified experimentally. In this perspective, the constraint we have found for the parameter $q$ should be interpreted as the values one would obtain {\it if} the gravitational baryogenesis is the actual mechanism to generate baryon asymmetry.  With these premises, it is, in our opinion, remarkable that the gravitational baryogenesis for the MEMe model can return such interesting values of its parameters. This result has profound implications for the theoretical foundations of generalized coupling theories and will be investigated in forthcoming works.

\section*{Aknowledgments} This work has been carried out in the framework of activities of the INFN Research Project QGSKY.

\end{document}